\documentclass[aps,prb,twocolumn]{revtex4-2}
\usepackage{graphicx,color}
\usepackage{hyperref}
\usepackage{amsmath} 
\newcommand{\bra}[1]{\langle #1\rangle}

\usepackage[normalem]{ulem}
\newcommand{\be}{\begin{equation}} 
\newcommand{\ee}{\end{equation}} 
\newcommand{\bea}{\begin{eqnarray}} 
\newcommand{\eea}{\end{eqnarray}}

\begin{document}
\title{Geometric percolation of spins and spin-dipoles in Ashkin-Teller model}
\author{ Aikya Banerjee, Priyajit Jana and P. K. Mohanty}
\email{pkmohanty@iiserkol.ac.in}
\affiliation {Department of Physical Sciences, Indian Institute of Science Education and Research Kolkata, Mohanpur, 741246 India.}
\date{\today}

\begin{abstract}
Ashkin-Teller model is a two-layer lattice model where spins in each layer interact ferromagnetically with strength $J$, and the spin-dipoles (product of spins) interact with neighbors with strength $\lambda.$ The model exhibits simultaneous magnetic and electric transitions along a self-dual line on the $\lambda$-$J$ plane with continuously varying critical exponents. In this article, we investigate the percolation of geometric clusters of spins and spin-dipoles denoted respectively as magnetic and electric clusters. We find that the largest cluster in both cases becomes macroscopic in size and spans the lattice when interaction exceeds a critical threshold given by the same self-dual line where magnetic and electric transitions occur. The fractal dimension of the critical spanning clusters is related to order parameter exponent $\beta_{m,e}$ as $D_{m,e}=d-\frac{5}{12}\frac{\beta_{m,e}}\nu,$ where $d=2$ is the spatial dimension and $\nu$ is the correlation length exponent. This relation determines all other percolation exponents and their variation wrt $\lambda.$ We show that for magnetic Percolation, the Binder cumulant, as a function of $\xi_2/L$ with $\xi_2$ being the second-moment correlation length, remains invariant all along the critical line and matches with that of the spin-percolation in the usual Ising model.
%For the electric percolation too the function remains invariant forming a new superuniversality class of percolation transition. 
 The function also remains invariant for the electric percolation, forming a new superuniversality class of percolation transition.
\end{abstract}

\maketitle

\section{Introduction}
Percolation theory is a fundamental concept in statistical physics \cite{Stauffer_1979,Essam_1980,Stauffer_1992Book,Sahimi_1994}. Although the concept was initially developed for purely applied problems, especially fluid flow in porous media \cite{Hammersley_1957}, it now finds application in various disciplines \cite{Sahimi_1994, Phillips_2006Book, Saberi_2015}. The phenomenology of percolation is intuitively simple but possesses a rich mathematical structure \cite{Kesten_1982, Grimmett_2013}. The fundamental percolation model involves randomly occupying sites or bonds on a regular lattice with probability $p$. A set of occupied sites connected by nearest neighbors is known as a connected cluster.
In an infinite lattice system, as the occupation probability $p$ increases, the largest cluster of connected sites or bonds grows progressively larger. At a specific threshold $p = p_c$, this cluster becomes extensive, spanning the lattice and creating connectivity across the system from one side to the other. This phenomenon represents a \textit{geometric phase transition}, a continuous transition driven solely by spatial arrangement and probabilistic occupancy \cite{Stauffer_1992Book}.
In a two-dimensional square lattice, the critical probability \( p_c \) for bond percolation is $\frac{1}{2} $, whereas for site percolation it is approximately \( 0.59274621 \) \cite{Ziff_2000}. Despite these differing threshold values, both bond and site percolation display identical critical properties, falling under the same \textit{universality class} known as ordinary percolation (OP). Percolation on lattices with various geometries also share this OP universality class. Research on percolation extends to higher dimensions, where similar principles and universality have been observed \cite{Stauffer_1992Book, Grimmett_2013}.

In ordinary site percolation sites are uncorrelated as they are occupied randomly and independently. When occupied sites are power-law correlated, the critical behavior of the system changes \cite{Zierenberg_PRE2017}.
Since long-range correlations are generated at critical points, one looks for percolation behavior there. In the ferromagnetic Ising system, neighboring spins prefer to have the same spin values, forming geometric clusters of parallel spins.  It is well known that these geometric clusters, both in $d=2$ and  $d=3$ \cite{Fortunato_2002, Saberi_2010} span the lattice when the temperature is lowered below a threshold value, that matches with the Curie temperature.  Thus, the percolation and magnetization transition occur together, but their critical behavior and universality classes are different, primarily because the exponents of the percolation transition can not be determined solely in terms of the Ising exponents. The study of correlated percolation has been further extended to Potts models.

In this article, we study site percolation transition in the Ashkin-Teller (AT) model \cite{Teller_1943}, which is a two-layer lattice model having ferromagnetic interaction $J$ of Ising spins in individual layers and a four-spin inter-layer interaction $\lambda$. The four-spin interaction is similar to an Ising-like interaction of spin-dipoles (product of spins in two layers). While spins lead to Magnetic ordering, spin-dipoles can produce an electric order or Polarization. In the isotropic AT model, both magnetic and electric transitions occur together along the critical line on the $\lambda$-$J$ plane, known as the Baxter line; the critical exponents of both transitions vary continuously. For the study of percolation, we need to consider three different kinds of clusters - the magnetic clusters formed by spins in individual layers and the electric cluster of spin dipoles. The first question is whether magnetic and electric percolation occur together. If so, whether they occur along the Baxter line? How do the critical exponents vary with $\lambda?$ What is the universality class of these correlated percolation transitions?

We find that the percolation transitions do occur along the Baxter line. Magnetic percolation follows the weak-universality hypothesis, where exponents vary, but their ratios remain invariant, and electric percolation violates both universality and weak-universality. On the other hand, they all follow a superuniversality hypothesis proposed recently \cite{Indranil_PRB2023}.

\subsection{Ordinary percolation}
The critical behavior of ordinary percolation transitions is characterized by a set of critical exponents. Since the critical exponents are related by scaling relations, in numerical simulations, one finds one of the two sets of critical exponents defined below.

{\it (a) Cluster exponents $\tau$, $\sigma$, $D$:} The distribution of the size of all connected clusters shows a characteristic scaling near the critical point. If $P(s)$ is the probability that a randomly chosen site belongs to a cluster of size $s$, then $P(s) = s^{-\tau}f(\epsilon s^\sigma)$ where $\epsilon = (p_c - p)$. At the critical point, the fractal dimension of the largest cluster is expressed as $D$. In two dimensions, the exponents are the following \cite{Stauffer_1992Book}.
\be
\tau = \frac{187}{91};~ \sigma = \frac{36}{91};~ D = \frac{91}{48}
\label{eq:clust_stand_pero}
\ee
{\it (b) Exponents $\nu$, $\beta$, $\gamma$:}
%(related to the correlation length, order parameter, and susceptibility ($\nu$, $\beta$, $\gamma$)} 
Another standard practice is to study the density of the largest cluster ($s_{max}$). The scaling of the order parameter $\phi_{s} = \langle s_{max} \rangle$ and the corresponding susceptibility ($\chi_{s}$) near the critical point gives the exponents $\beta$ and $\gamma$, respectively. In particular, $\phi_{s} \sim |\epsilon|^{\beta}$ and $\chi_{s} \sim |\epsilon|^{-\gamma}$. The critical exponent $\nu$ is associated with the scaling of the correlation length ($\xi$), $\xi\sim |\epsilon|^{-\nu}$. In two dimensions, the exponents are,
\be 
\nu= \frac43;~ \beta = \frac{5}{36};~ \gamma= \frac{43}{18}.
\ee
These exponents are not independent; they relate to the cluster-exponents in Eq. \eqref{eq:clust_stand_pero} by the following scaling relations \cite{Stanley_Book_1971, Stauffer_1979}.
\be
\tau = 2 + \frac{\beta}{\beta + \gamma};~ \sigma = \frac{1}{\beta + \gamma};~ D = d - \frac{\beta}{\nu},
\ee
where $d=2$ is the dimension of the system.

The two sets of exponents reflect distinct aspects of percolation transition. While the cluster exponents reflect on the fractal nature of the clusters and their scaling properties, the other set of exponents ($\nu, \beta, \gamma$) features how connectivity appears in large scales, leading to long-range order in the system.

However, this notion of a geometric phase transition is not confined to simple, randomly occupied lattices. The percolation framework has been expanded to explore connectivity and phase transitions in complex networks, fractal graphs, and systems spanning various dimensions \cite{Grassberger_1986}. The study of percolation in correlated systems has emerged as one of the most significant extensions of this concept, revealing new facets of critical behavior.

%%%%%%%%%%%%%%%%%%%%%%%%%%%%%%
\subsection{Correlated percolation}
%%%%%%%%%%%%%%%%%%%%%%%%%%%%%%
Ordinary percolation generally deals with the emergence of global connectivity in a system with randomly distributed constituents dictated by geometry and probability. However, percolation can be studied in systems with inherent correlation among constituents \cite{Abel_PRB_1984, Zierenberg_PRE2017}. In the last few decades, the percolation transition has been extensively studied in the context of the two-dimensional Ising model. In particular, Fortuin and Kasteleyn (FK)studied a random cluster model (RCM) \cite{Fortuin_1972}, similar to the Ising model on a two-dimensional square lattice except that the neighboring spins interact with strength $J = \infty$ with probability $p$ and $J = 0$ with probability $(1-p)$. The model exhibits a phase transition of percolation at $p_c = 1- e^{-2/k_B T_c}\simeq 0.585786$, where $T_c = 2 / \ln(1+\sqrt{2})$ is the critical temperature associated with the ferromagnet-to-paramagnet transition in the Ising model. Clusters are formed by joining the neighboring sites with parallel spins. Using standard percolation techniques, one can, in practice, estimate the critical exponents. The critical behavior in the FK clusters belongs to the Ising universality class (IUC) with the same critical exponents.
\be 
\nu=1;~ \beta = \frac{1}{8};~ \gamma= \frac74
\label{eq:Ising}
\ee
Corresponding percolation cluster exponents are $\tau= \frac{31}{15}$, $ \sigma=\frac{8}{15}$,$D=\frac{15}8.$

Another approach is to study the geometric clusters of the usual Ising model with isotropic ferromagnetic interaction $J$ among neighboring spins. The geometric clusters are
constructed by joining neighboring spins when they are parallel. It turns out that in a steady state configuration, the cluster that contains maximum number of sites, say $s_{max}$ sites, becomes macroscopic (carrying a finite fraction of total $L^2$ sites) when the temperature $T$ is less than the Onsager value $T_c = 2J/ \ln(1+\sqrt{2})$ \cite{Onsager_1944}.
Although $T_c$ is identical to that of the ferromagnetic transition, the critical behavior of this geometric percolation is different from IUC.
It was argued by Stella and Vanderzande \cite{Stella_1989} that the exponents of the geometric percolation can not be expressed solely in terms of the Ising exponents. They conjectured that the fractal dimension of the percolating cluster at the critical point is not $D=d- \beta/\nu= 15/8,$ 
rather it is
\be
D^P = d- w \frac{\beta}{\nu}= \frac{187}{96};~~ w = \frac{5}{12},
\label{eq:dfP}
\ee
where the additional parameter $w$ is determined from the connection of Ising percolation with tri-critical $q=1$ Potts model \cite{Stella_1989}.
Since the percolation transition occurs at the same critical temperature, the correlation length $\xi$ of the system, and thus the corresponding exponent $\nu$ remains invariant.
This allows us to write $D^P = d- \frac{\beta^P}{\nu^P}$ where,

\bea
%D^P = d- \frac{\beta^P}{\nu^P}; ~~
\nu^P= \nu=1;~~\beta^P= \frac{5\beta}{12}=\frac{5}{96};~~\gamma^P=\frac{13\gamma}{12}=\frac{91}{48}.
\label{eq:Z2P}
\eea
Here, the superscript $P$ stands for exponents related to the percolation transition, and in the last step, we use the scaling relation, $\gamma^P = d\nu^P-\beta^P.$

Indeed, geometric percolation transition of spins in the Ising model forms a new universality class where $\phi = \bra{s_{max}}/L^2$ plays the role of order parameter leading to critical exponents \eqref{eq:Z2P};  we refer to this universality class as $Z_2$P (percolation associated to $Z_2$ symmetry breaking, that is Ising transition). Not only the critical exponents but the scaling functions of $Z_2$P are also different from the corresponding scaling function of IUC. It was shown by Coniglio and Klein \cite{Coniglio_1980, Coniglio_2001} that when one takes the geometric clusters and deletes the connected bonds with probability $p = 1 - e^{-2J/k_BT}$ the fractal dimension of critical spanning cluster reduces and the percolation transition of these new clusters, formally known as the Fortuin-Kasteleyn Coniglio-Klein (FKCK) clusters \cite{Fortuin_1972,Coniglio_1980} belong to IUC.

\subsection{Aim of this work}
In this article,
we investigate the percolation transition in the Ashkin-Teller (AT) model, a system known to exhibit continuous variation of critical exponents \cite{Wu_1974}. The model is defined on a bilayer square lattice with Ising spins $\sigma$ in one layer and $\tau$ on the other.  Along with the usual intra-layer ferromagnetic Ising interaction of strength $J$, the interaction of spin dipoles $\alpha_{\bf i}= \tau_{\bf i} \sigma_{\bf i}$ with their neighbors with strength $\lambda$
(as shown in Eq. \eqref{hamiltonian} below) brings in nontrivial critical features to the system. AT model shows a phase transition from being an un-polarized paramagnet to a polarized ferromagnet. This transition occurs on a critical line on the $\lambda$-$J$ plane, formally known as the Baxter line (see Fig. \ref{fig:phaseAT}). The critical exponents of both the transitions depend explicitly on the intra-layer interaction strength $\lambda;$ the functional form of variations are known exactly from the mapping of the model to eight vertex (8V) model \cite{Baxter_1972}. For magnetic transition, although the exponents $\beta, \gamma, \nu$ depend on $\lambda$, the ratios $\frac\beta\nu,$ $\frac\gamma\nu$ remain invariant, for electric (Polarization) transition, both the exponents and their ratios vary continuously. The dependence of critical exponents on system parameters poses a threat to the universality hypothesis. Later, it was realized that the existence of a marginal parameter (like the four spin interaction term in AT model \cite{Kadanoff_AnPhys1979}) can indeed generate a continuous variation of the critical exponents.

 A general question is then: what carries the universal signature of the underlying critical point (governed by the relevant operators) when critical exponents vary along a marginal direction? Recently, it was proposed \cite{Bonati_PRL2019, Indranil_PRB2023} that Binder cumulant as a function of $\xi_2/L,$ $\xi_2$ being the second-moment correlation length, is a superuniversal function which remains invariant along the critical line formed by the marginal operator, even though the critical exponents vary.
 This function for magnetic transition is identical to that of Ising universality ($Z_2$), and for the electric transition, it matched with that of the Polarization transition of the bilayer Ising model, which we refer to as $Z_2^2$ (the superscript $2$ refers to spin dipoles in {\it two} independent Ising systems).

We aim to study geometric percolation transition on (along) the Baxter line. Since for $\lambda=0,$ the two layers decouple, the usual Ising transition occurs independently on both the layers and thus, the magnetic percolation transition there belongs to the $Z_2$P universality class with exponents \eqref{eq:Z2P}. In particular, the fractal dimension of the spanning cluster there is $D_m=d- w\frac{\beta_m}\nu,$ with magnetization exponents $\beta_m=1/8,\nu=1$, and $w=\frac{5}{12}.$ Since $\beta_m$ and $\nu$ both change with $\lambda$ we would like to investigate whether the fractal dimension $D_m$ also varies with $\lambda.$ In general, a question of importance is the fate of $w$ - will it vary with $\lambda?$ The same questions can also be asked about $D_e,$ the fractal dimension of the polarization clusters.

 Since transitions along a critical line extended by a marginal operator form a unique superuniversality class,
one expects the respective percolation transitions to also obey superuniversality. Any dependence of $w$ on $\lambda$ would alter the functional relation among respective percolation exponents. We conjecture that $w$ remains invariant along the Baxter line and it must be $w=\frac{5}{12}$ (which is the value for $\lambda=0$);
in other words,
the fractal dimension critical percolation clusters must be,
\be 
D_m=d- \frac{5}{12}\frac{\beta_m(\lambda)}{\nu(\lambda)};~ D_e=d- \frac{5}{12}\frac{\beta_e(\lambda)}{\nu(\lambda)}. \label{eq:conj}
\ee
If this conjecture turns out to be true, can fix all other critical exponents of percolation transitions (see Eq. \eqref{eq:exact_expP} below).

We verified this conjecture using extensive simulations of the AT model. The results can be summarized as follows. First, the Baxter line remains critical for both magnetic and electric percolation transitions; the critical exponents of the transitions vary continuously along the critical line following the conjecture \eqref{eq:conj}. Like magnetization transition, the magnetic percolation too obeys the weak universality hypothesis \cite{Suzuki_1974} where the ratio of exponents remains invariant. In electric percolation, however, all the critical exponents and their ratios vary with $\lambda.$

In addition, we find that the
percolation transition in the AT model is described by three different superuniversal functions corresponding to three different universal behaviors: (a) $Z_2$P that describes correlated percolation of spins, (b) $Z_2^2$P which describes the percolation of spin-dipoles and
(c) $Z_4$P corresponds to percolation transition at the special point $\lambda=\ln(3)/4$ where the model is equivalent to 4-state Potts model.

The article is organized in the following way. In Section II, we briefly describe the model along with the phases in the model (II-A) and associated transitions (II-B). This is followed by an overview of the numerical scheme (Section II-C), an analysis of the model's critical behavior (Section II-D), and a study of cluster properties (Section II-E). Next, we describe the percolation exponents (Section II-F). In addition, we highlight the properties of percolation at a special point in the model, where the phase transition is governed by $Z_4$ symmetry breaking (Section III). In Section IV, we introduce the concept of superuniversality, which connects systems with continuously varying critical exponents through invariant scaling behavior. Finally, in Section V, we summarize the findings and discuss their implications.

\section{The model}
The Ashkin-Teller (AT) model \cite{Teller_1943} is a two-layer lattice model, where Ising spins in each layer, referred to as $\sigma$ and $\tau$, interact ferromagnetically, and among different layers they interact via the four-spin coupling term. On a periodic $L\times L$ square lattice, the spins at site ${\bf i} \equiv (x,y)$ where $x,y= 1,2,\dots L$ are denoted as $\sigma_{\bf i}=\pm 1$ and $\tau_{\bf i}=\pm 1.$ In isotropic case, the energy of any given configuration $\{\sigma_{\bf i},\tau_{\bf i}\},$ is given by a Hamiltonian,
\begin{equation} 
\mathcal{H}= - J\sum_{\bra{{\bf i},{\bf j}}} \left( \sigma_{\bf i} \sigma_{\bf j} +\tau_{\bf i} \tau_{\bf j}\right) - \lambda \sum_{\bra{{\bf i},{\bf j}}} \alpha_{\bf i} \alpha_{\bf j}; ~~ \alpha_{\bf i} = \sigma_{\bf i}\tau_{\bf i}
\label{hamiltonian}
\end{equation}
where the sum is taken over nearest neighbor pairs $\bra{{\bf i},{\bf j}}$. The first two terms correspond to intra-layer spin interactions similar to that of the  standard Ising model, and we assume $J>0$ (ferromagnetic interaction). The last term introduces an interaction of spins in different layers with strength $\lambda.$ 
We refer to the product of spins at site ${\bf i}$ as the spin dipoles $\alpha_{\bf i}=\pm1$; the values $\pm1$ may be considered as the orientation of the spin-dipole at ${\bf i}.$ 
The inter-layer interaction is an Ising-interaction of dipoles, and thus, depending on the value of $\lambda$, the system can lead to a ferromagnetic (anti-ferromagnetic) dipolar-order or Polarization for $\lambda>0$ ($\lambda<0$) respectively even when individual spins $\sigma_{\bf i}, \tau_{\bf i}$ are in paramagnetic state.

%%%%%%%%%%%%%%%%%%%%%%%%%%%%%%%%%%%%%%%%%%%%%%
\subsection{Phases and phase transitions}
\begin{figure}
 \centering
 \includegraphics[width=0.9\linewidth]{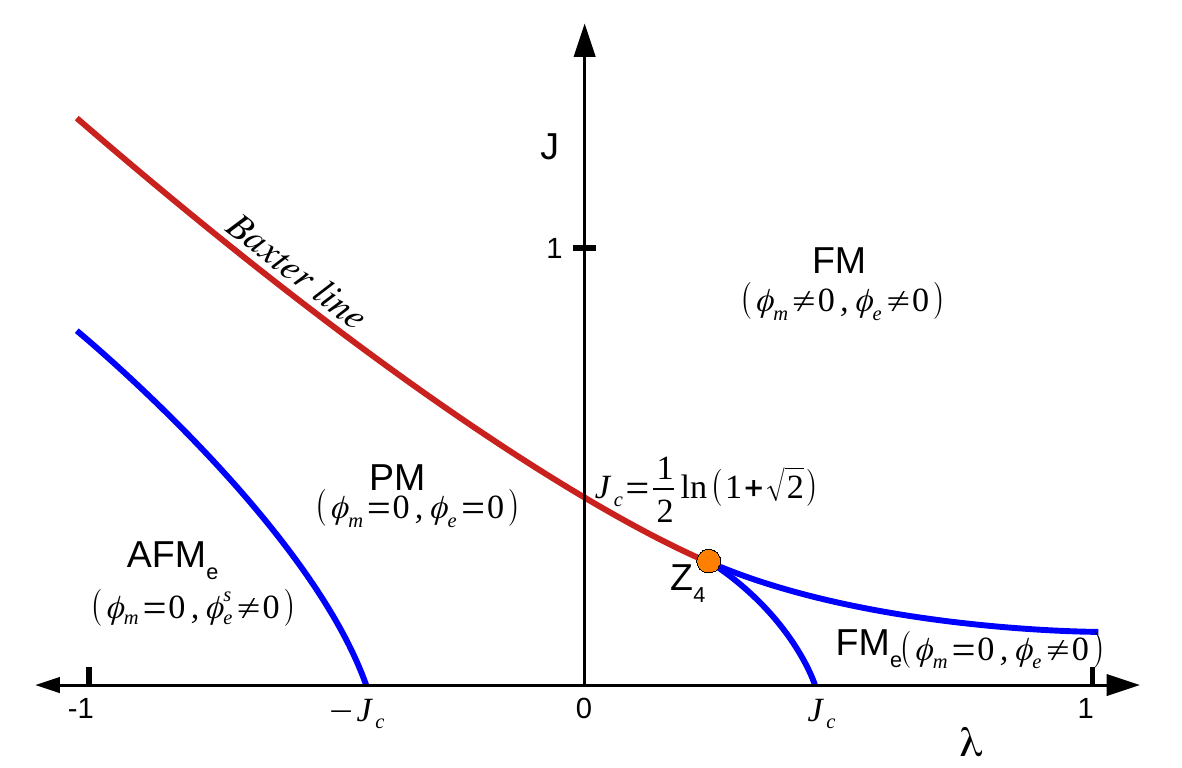}
 \caption{(Color online) Four phases of the Ashkin-Teller model in $\lambda$-$J$ plane for $T=1$: Unpolarized paramagnet (PM) where neither spins $\tau,\sigma$ nor spin-dipoles $\alpha =\tau\sigma$ orders, polarized ferromagnet (FM) where both orders and two other paramagnetic states where spin dipoles order ferromagnetically (FM$_e$) or anti-ferromagnetically (AFM$_e$). The Baxter line $J = \frac12\sinh^{-1}(e^{-2 \lambda})$ extends in the range $\lambda<\lambda^*= \ln(3)/4$ and separates PM from FM (marked in red). }
 \label{fig:phaseAT}
\end{figure}

%%%%%%%%%%%%%%%%%%%%%%%%%%%%%%%%%%%%%%%%%%%%%%
The obvious symmetries of the system are: $\sigma \rightarrow-\sigma$, $\tau\rightarrow -\tau$ and $\sigma\rightarrow \tau$. These symmetries allow $\phi_m=\bra{\sigma_{\bf i}}=\bra{\tau_{\bf i}}$ and $\phi_e=\bra{\sigma_{\bf i}\tau_{\bf i}}\equiv \bra{\alpha_{\bf i}}$ as independent order parameters that describe the magnetic and electric order respectively. Accordingly, the AT model exhibits four phases, shown in Fig. \ref{fig:phaseAT}: a ferromagnetic phase (FM) where both $\phi_m$ and $\phi_e$ are nonzero, a pure paramagnetic (PM) phase where $\phi_m=0=\phi_e,$ and two polarized paramagnetic phases where $\phi_m=0$ but the spin dipoles $\alpha_{\bf i}$ exhibits a ferromagnetic or antiferromagnetic electric order, referred to as FM$_e$, and AFM$_e$ respectively. In the AFM$_e$ phase $\phi_m=0=\phi_e,$ but a staggered electric order exists on each of the sub-lattices, i.e., $\phi^s_e = \bra{ (-1)^{\bf i. q} \alpha_{{\bf i}} } \ne 0$ where vector ${\bf q }\equiv (1,1)$.  Different lines mark the transition between these phases. The line marked as the Baxter line is of particular interest; when $J$ or $\lambda$ or both are increased beyond this line, the system acquires both magnetic and electric orders simultaneously. Equation of
Baxter line,
\be
\sinh(2\beta J) = e^{-2 \lambda}
\label{eq:BL}
\ee
is known exactly from the exact mapping \cite{Wu_PRB_1970, Kadanoff_PRB_1971} of AT model and eight vertex (8V) model \cite{Baxter_1972}. This mapping also provides exact critical exponents \cite{Wu_1974, Domany_1979} of both magnetic and electric transitions, which vary continuously with $\lambda,$ as one moves along the Baxter line. The reason for this continuous variation owes to the fact that the inter-layer coupling parameter $\lambda$ is marginal \cite{Kadanoff_AnPhys1979} (does not flow under renormalization group operations).

For completeness, let us describe the critical behavior explicitly using $\Delta= T_c-T$ as the control parameter, where $T_c=1$ on the Baxter line. For a thermodynamically large system, the order parameter $\phi_{m,e}$ and the susceptibility $\chi_{m,e}$ near $T=T_c$ scales as,
\begin{align}
 \phi_{m,e} \sim|\Delta|^{\beta_{m,e}}, \quad\chi_{m,e} 
 \sim |\Delta|^{\gamma_{m,e}}.
\end{align}
Moreover, the correlation function $C(r)$ defined as,
\begin{align*}
 &C(r) = \bra{\sigma_{\mathbf{i}}\sigma_{\mathbf{i+r}}} - \phi_{m}^2 = \bra{\tau_{\mathbf{i}}\tau_{\mathbf{i+r}}} - \phi_{m}^2. 
% &C_{e}(r) = \bra{\alpha_{\mathbf{i}}\alpha_{\mathbf{i+r}}} - \phi_{e}^2
\end{align*}
Away from the critical point, the correlation function decays exponentially fast when $r= |{\bf r}|$ is large: $C(r)\sim e^{-r/\xi}$ 
where $\xi$ is the correlation length. It diverges as one approaches the critical point as $\xi \sim \Delta^{-\nu}$, giving rise to another exponent $\nu.$ Note that $\xi$ is an emergent length scale of the system, and more often than not, it is independent of the operator whose correlation is measured. Thus, $\xi$ must be the same for both electric and magnetic transition, and $\nu$ does not carry the subscripts $m,e.$ From the correspondence of AT model and 8V model, the the exponents are known exactly \cite{Wu_1974, Domany_1979},
\bea
&\nu = \frac{2(\mu - \pi)}{(4\mu - 3\pi)}, \text{ where } \cos\mu = e^{2\lambda}\sinh{(2\lambda)};\cr
&\beta_m = \frac{\nu}{8}, \gamma_m = \frac{7\nu}{4}; \quad \beta_e = \frac{2\nu-1}{4}, \gamma_e = \frac{1}{2} + \nu.
\label{eq:exact_exp}
\eea
Moreover, although the critical exponents vary continuously with $\lambda$, they satisfy the hyper-scaling relations everywhere,
\begin{equation}
 2\beta_{m,e} + \gamma_{m,e} = d \nu %= \frac{d \gamma_{m,e}}{2-\eta_{m,e}}.
 \label{eq:scaling-I}
\end{equation}
Here $d=2$ is the dimension of the system.

Some special cases of the model need special mention. For $\lambda=0,$ the spins $\tau$ and $\sigma$ are decoupled, and thus, the magnetic critical exponents must reduce to the Ising exponents in Eq. \eqref{eq:Ising}, which carries the signature of $Z_2$ symmetry breaking. The spin dipoles also undergo a phase transition at $\lambda=0.$ 
Although $\tau$ and $\sigma$ are independent, in the supercritical phase, both produce macroscopic regions of parallel spins in both layers, which must have a large spatial overlap; thus, their product $\alpha= \tau\sigma$ also acquires the same value in the overlapping region producing finite polarization. Accordingly
the the electric transition here must have
critical exponents
\be
 \nu=1;~ \beta_e = \frac{1}{4};~ \gamma_e= \frac32,
\label{eq:Z22}
\ee
where $\nu$ remains same, $\beta$ is doubled and $\gamma_e= d- 2 \beta_e.$
We refer to this critical behavior as $Z_2^2$ universality class; the superscript reminds us of two layers of broken the $Z_2$ symmetry.

Another special case is $\lambda=J.$ Here, the model is invariant under the permutations of the four states $\{\sigma=\pm, \tau=\pm\})$
and thus the critical behavior here must be in the universality class of $q=4$ Potts model that breaks $Z_4$ symmetry
with exponents \cite{Enting_1975, Wu_Potts_RMP_1982}
\be
 \nu_4=\frac43;~ \beta_4 = \frac{1}{12} ;~ \gamma_4= \frac76
 \label{eq:Z4exp}
\ee
Here, the subscript $4$ is used to remind us about the $Z_4$ universality class. At $Z_4$ point on Baxter line we get $\nu= \nu_4,$ $\beta_e=\beta_m= \beta_4$ and $\gamma_e=\gamma_m= \gamma_4.$

Although the critical exponents of AT model vary continuously they satisfy additional relations. The magnetic exponents obey the well-known weak-universality criterion \cite{Suzuki_1974}, i.e., exponents $\beta_m, \gamma_m, \nu$ vary continuously as $\lambda$ varies, but their ratio $\beta_m/\nu$ and $\gamma_m/\nu$ remains same as that of the Ising universality class. On the other hand, electric exponents are believed to be non-universal \cite{Samaj_PRE2018}. Recently, however, it has been shown \cite{Indranil_PRB2023} that both the transitions form a superuniversality class in the sense that a superuniversal function,
namely Binder-cumulants as a function of $\frac{\xi_2}{L},$ remain invariant all along the Baxter line even when the critical exponents vary. Here, $\xi_2$ is the second-moment correlation length, defined for magnetic and electric transitions as,
\be \xi_2 = \left[ \frac{\sum_{\mathbf{r}} r^2 C(\mathbf{r})}{\sum_{\mathbf{r}}C(\mathbf{r})}\right]^{\frac12}
\label{eq:xi2}
\ee
This function for magnetic transition along the Baxter line is the same as that of the ordinary Ising model, and thus, the transition belongs to the $Z_2$-superuniversality class. The function for electric transition along the critical line is unique even when all exponents vary with $\lambda$, and we refer to this transition form a new superuniversality class $Z_2^2,$ where the superscript reminds us that the transition is wrt spin-dipoles.

%%%%%%%%%%%%%%%%%%%%%%%%%%%%%%%%%%%%%%%%%%%%%%%%%%%
\subsection{Magnetic and electric percolation transitions}
%%%%%%%%%%%%%%%%%%%%%%%%%%%%%%%%%%%%%%%%%%%%%%%%

\begin{figure}
 \centering
 \includegraphics[width=\linewidth]{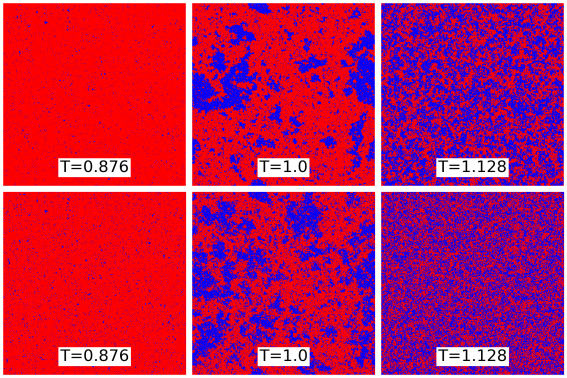}
 \caption{(Color online) Typical steady state configurations of $\tau$-spin(top row) and spin-dipoles $\alpha$ (bottom row) at super-critical($T<T_c$), critical ($T=T_c=1$) and sub-critical ($T>T_c$) states for $L=512,$ $\lambda=0.1$ on Baxter line.}
 \label{fig:transition_in_pic}
\end{figure}
In this article, we aim to study the percolation properties of the AT model. There are three different kinds of clusters here: geometric clusters of spins in two different layers, which are equivalent due to isotropy, and clusters formed by the spin-dipoles. The clusters are formed in the same way one defines clusters of site percolation: if neighboring spins on a layer are parallel, they belong to the same magnetic cluster, whereas the same dipolar orientation in neighboring sites belongs to a unique electric cluster. For any configuration $C\equiv (\{ \sigma_{\bf i}\}, \{ \tau_{\bf i}\})$ if the number of  clusters of
spins $\sigma$ and $\tau$ and that of spin dipole
type $\alpha=\tau\sigma$ are respectively $K_{\sigma,\tau,\alpha}$ and if $s^k_{\sigma,\tau,\alpha}$ denotes the size or mass of $k$-th cluster (i.e., number of lattice sites belonging to $k$-th cluster) then we must have
\be L^2 = \sum_{k=1}^{K_\sigma} s^k_\sigma = \sum_{k=1}^{K_\tau} s^k_\tau =\sum_{k=1}^{K_\alpha} s^k_\alpha.
\ee

\begin{figure}
 \includegraphics[width = 0.485\linewidth]{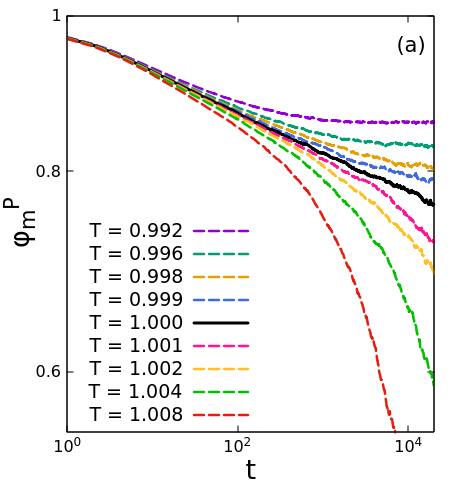}
 \includegraphics[width = 0.485\linewidth]{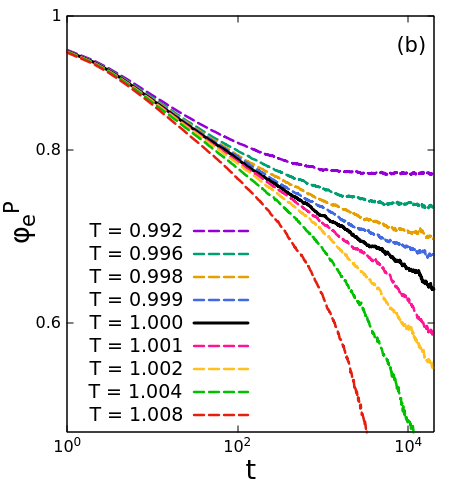}
 \includegraphics[width = 0.485\linewidth]{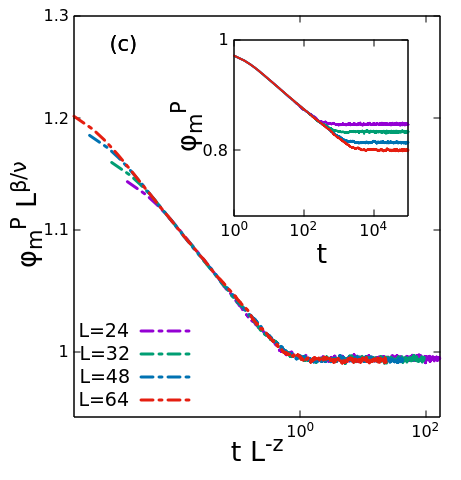}
 \includegraphics[width = 0.485\linewidth]{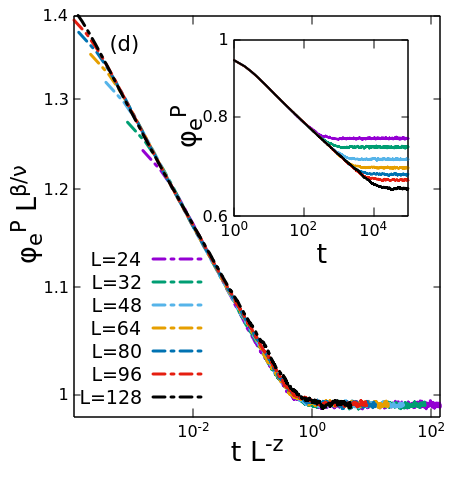} 
 \caption{(Color online) (a) \& (b) reflects power-law decay of $\phi_m^P$ and $\phi_e^P$ at critical temperature $T_c = 1$ (system size, $L = 256$ and $\lambda = 0.1$). Compensated time plots for different L's are shown in (c) \& (d) with uncollapsed plots in the inset. Here, we average over $10^4$ independent runs from the ordered state.}
 \label{fig:t}
\end{figure}
For every configuration $(\{ \sigma_{\bf i}\}, \{ \tau_{\bf i}\})$ one among $K_{\sigma,\tau,\alpha}$ clusters is the largest and we denote its size as $s^{\sigma,\tau,\alpha}_{max}.$
In equilibrium, we must have $\bra{s^\sigma_{max}} =
\bra{s^\tau_{max}}$ and one can consider $\phi^P_m= \frac{1}{L^2} \bra{s^\tau_{max}}$ as the order parameter that characterizes the magnetic percolation transition.
Similarly, $\phi^P_e= \frac{1}{L^2} \bra{s^\alpha_{max}}$ is the order parameter for the magnetic percolation transition. When $(J, \lambda)$ are taken on the Baxter line, the critical temperature is $T_c=1$ and one expects that both $s^{\tau}_{max}$ and $s^{\alpha}_{max}$ becomes macroscopic in size (finite fraction of $L^2$ site belongs to this largest cluster) when temperature $T$ is lowered below $T_c=1.$ For $\lambda=0.1,$ the typical configurations of magnetic and electric clusters near $T=T_c$ are shown in Fig. \ref{fig:transition_in_pic}.

The critical exponents of these percolation transitions are defined near $T=T_c$ in terms of $\Delta= T_c-T$ as,
\begin{eqnarray}
 && \phi_{m}^P \sim \Delta^{\beta_{m}^{P}}; 
 \chi^P_{m} = \frac{ \bra{(s_{max}^{\tau})^2} - \bra{s_{max}^{\tau}}^2}{L^4} \sim \Delta^{\gamma_{m}^{P}}\cr 
 &&\phi_{e}^P \sim \Delta^{\beta_{e}^{P}}; 
 \chi^P_{e} = \frac{ \bra{(s_{max}^{\alpha})^2} - \bra{s_{max}^{\alpha}}^2}{L^4} \sim \Delta^{\gamma_{e}^{P}}.
\end{eqnarray}
As usual, the critical exponent $\nu,$ associated with the emergent length scale of the system is not altered. In the following, we estimate these critical exponents from Montecarlo simulation, using finite size scaling (FSS) analysis \cite{Stanley_Book_1971, Binder_FSS_1981}.

\subsection{Simulation and dynamics}
We perform Monte Carlo simulation of the model on a $L\times L$ system using Glauber dynamics, where a randomly chosen spin ($\sigma$ or $\tau$) is flipped with Metropolis rate $r={\rm Min}(1,e^{-\beta\Delta E})$ where $\beta = 1/k_B T$ and $\Delta E$ measures the energy difference; $r=1$ when energy of the target configuration lower.
The study is conducted along the Baxter line, where for any given $\lambda$, we take $J=\sinh^{-1}( e^{-2\lambda})/2$  (from Eq. \eqref{eq:BL}) and vary temperature $T$.

First, we determine the critical temperature.
For a reasonably large system $L=256,$ starting from a fully magnetic state where $\sigma_{\bf i}=1= \tau_{\bf i}$ we evolve the system using Metropolis algorithm and measure the size of the largest magnetic cluster $s^\tau_{max}$ and largest eclectic cluster $s^\tau_{max}$ at time $t$ using Hoshen–Kopelman algorithm \cite{HK_Algo_PRB_1976}, which is then averaged over $10^4$ samples. The results for $\lambda=0.1,$ are shown in Fig. \ref{fig:t} in log-scale. Figure \ref{fig:t}(a),(b) shows evolution of $\phi_m^P(t)= \bra{\tau_{\bf i}}/L^2$ and $\phi_e^P(t)= \bra{\alpha_{\bf i}}/L^2$ respectively. For small $T,$ both order parameters saturate to a non-zero steady-state value, whereas for large $T$, they decay exponentially. At $T=T_c,$ one expects a power-law decay,
\be 
\phi_{m,e}^P(t) \sim t^{-\theta_{m,e}^P}
\ee
In fact, for a large system, $T_c$ can be determined as the
temperature for which $\phi_{m,e}^P(t)$ vs $t$ in log-scale becomes linear for large $t.$ We also obtain the decay exponents $\theta_{m,e}^P$ as the slope of this straight line. From Fig. \ref{fig:t} we find that $\theta_m^P= 0.02(6)$ and $\theta_e^P= 0.04(3).$
Again, when $T\ne T_c,$ the order parameters obey a scaling form, $\phi_{m,e}^P(t) = L^{-\theta_{m,e}z} h( t/L^z),$ where $z$ is the dynamical exponent of the system which relates the relaxation time to the size of the system. In Fig. \ref{fig:t}(c), (d) we plot $\phi_{m,e}^P(t) L^{\theta_{m,e}z} $ as a function of $t/L^z$ using $z$ as a fitting parameter. The best collapse is obtained when $z=2.0(0)$ for spin clusters and $z=2.0(5)$ for spin-dipole clusters. Note that, like the correlation length $\xi,$ the relaxation time $\xi_t$ is an emerging timescale of the system and does not depend on the choice of order parameter or any other observable of interest. Near the critical point $\xi_t \sim \Delta^{-\nu_t}$ where $\nu_t$ is the correlation time exponent and it is related to $z, \theta_{m,e}$ by the scaling relations
\be
z= \frac{\nu_t}{\nu}; ~~ \theta_{m,e} = \frac{\beta_{m,e}}{\nu_t}.
\ee
From estimates of the dynamic exponents $\theta_{m,e}$ and $z$ we expect to get $ \frac{\beta_{m,e}}{\nu} =z \theta_{m,e}$ ($= 0.05(2),0.08(8)$ for $\lambda=0.1$).
We will verify this in the next section, where we obtain independent estimates of $\beta_{m,e}$ and $\nu$ from the steady state data.

\subsection{Static exponents}
The static exponents are determined from finite size scaling of the steady state values of the Binder cumulants, the order parameters, and the susceptibilities near the critical point. The Binder cumulant is defined as \be B^P_{m,e} = 1-\frac{ \bra{(s_{max}^{\tau,\alpha})^4}}{ 3\bra{(s_{max}^{\tau,\alpha})^2}^2} = {\cal F}_{m,e}^B( \Delta L^{1/\nu}). \ee
Note that the argument of ${\cal F}^{m,e}_B$ is dimension less because in finite systems, the correlation length $\xi$ is limited by the system size $L$ and thus $L \equiv \xi \sim \Delta^{-1/\nu}.$ Similarly, the order parameter and the susceptibility for finite $L$ scales as,
\bea \phi^P_{m,e} = \Delta ^{\frac{\beta_{m,e}^P}{\nu}}{\cal F}_{m,e}^\phi( \Delta L^{1/\nu}); \chi^P_{m,e} = \Delta ^{-\frac{\gamma_{m,e}^P}{\nu}}{\cal F}_{m,e}^\chi( \Delta L^{1/\nu}). \nonumber
\eea

\begin{figure}
\centering
 \includegraphics[width = 0.485 \linewidth]{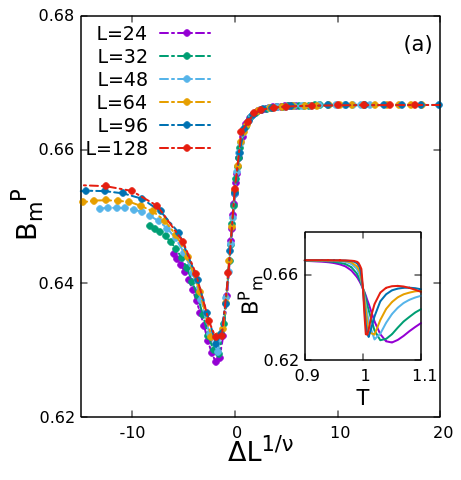}
 \includegraphics[width = 0.485 \linewidth]{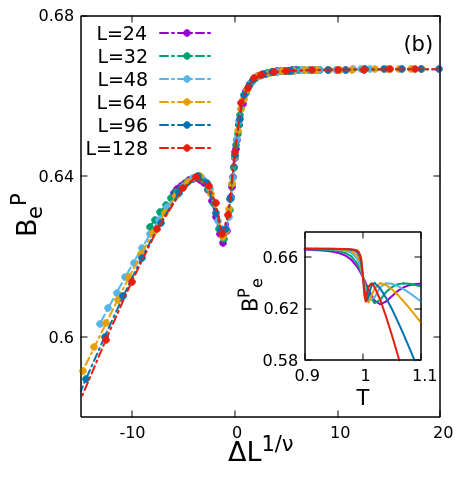} 
 \includegraphics[width = 0.485 \linewidth]{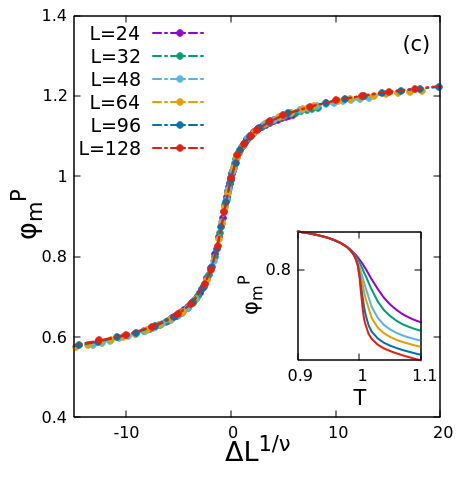}
 \includegraphics[width = 0.485 \linewidth]{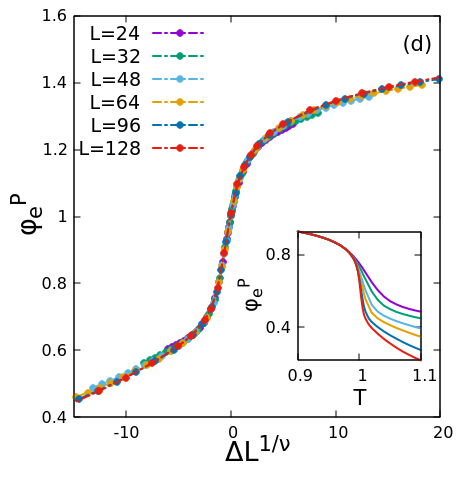}
 \includegraphics[width = 0.485 \linewidth]{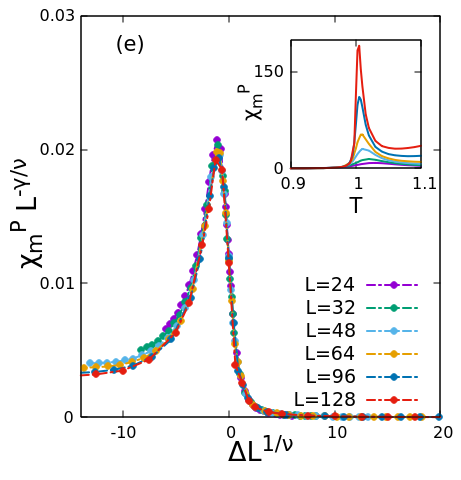}
 \includegraphics[width = 0.485 \linewidth]{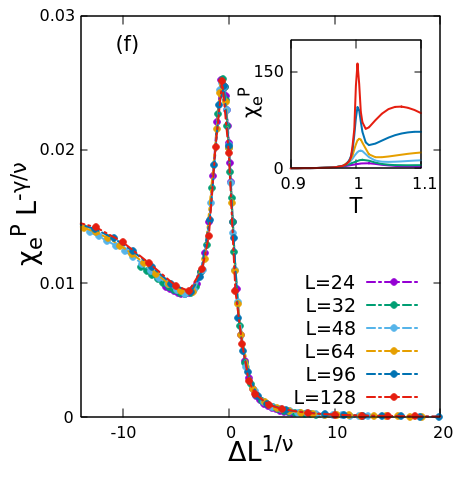}
 \caption{ (Color online) Data collapse obtained for $B_m^P$ (a), $B_e^P$ (b), $\phi_m^P$ (c), $\phi_e^P$ (d), $\chi_m^P$ (e) and $\chi_e^P$ (f) for $\lambda = 0.1$. The best collapse is obtained for $\nu = 0.87(8)$ (both magnetic and electric), $\frac{\beta_{m,e}^P}{\nu} = 0.052(1),0.09(0)$, $\frac{\gamma_{m,e}^P}{\nu} = 1.89(6), 1.80(9)$. The uncollapsed plots are shown in the inset. Data are averaged over $10^7$ or more samples in a steady state.}
 \label{fig:collapse}
\end{figure}

We use the following strategy to determine the critical exponents from FSS \cite{Binder_FSS_1981, Cardy_FSS_1996}. From the Monte Carlo simulations, we measure the steady state average
of $s_{max}^{\tau,\alpha}$ and its second and fourth moments as a function of temperature $T$ (for a fixed value of $\lambda, J$ on the Baxter line), and then repeat the calculation for different $L.$ First the critical temperature $T_c$ is determined from the crossing point of $B^P_{m,e}$ \cite{Binder_FSS_1981, Binder_Cum_PRL_1981} which help us to know $\Delta= T_c-T.$ From the plot of $B^P_{m,e}$ as a function of $\Delta L^{1/\nu}$ for different $L,$ using $1/\nu$ as a fitting parameter that gives best data collapse, we get the estimate of $\frac1\nu.$ This is shown for $\lambda=0.1,$ in Fig. \ref{fig:collapse} (a), (b) for both magnetic and electric cases; we find $1/\nu =1.13(9)$, which matches with the exact values $\nu=0.87(8)$ known from Eq. \eqref{eq:exact_exp}. This is an explicit check that the critical point and the correlation length exponents are not altered.

Now, we proceed to determine the order parameter and susceptibility exponents. It is clear from above scaling form that $\phi^P_{m} \Delta ^{-\frac{\beta_{m}^P}{\nu}}$ as a function of $\Delta L^{1/\nu}$ can be made to collapse using $\frac{\beta_{m}^P}{\nu}$ as the fitting parameter. From the best collapse, in Fig. \ref{fig:collapse} (c) we find $\frac{\beta_{m}^P}{\nu}=0.052(1).$ Figure \ref{fig:collapse} (d) describes the collapse of $\phi^P_{e} \Delta ^{-\frac{\beta_{e}^P}{\nu}}$ versus $\Delta L^{1/\nu}$ with $\frac{\beta_{e}^P}{\nu}= 0.09(0).$ In a similar way from the plot of $\chi^P_{m,e} \Delta ^{\frac{\gamma_{m,e}^P}{\nu}}$ as a function of $\Delta L^{1/\nu}$ for different $L$ we estimate that $\frac{\gamma_{m,e}^P}{\nu}=1.89(6),1.80(9)$ respectively. Form the values of $1/\nu,$ $\frac{\beta_{m,e}^P}{\nu}$ and $\frac{\gamma_{m,e}^P}{\nu}$ one can find the values of $\beta_{m,e}^P,\gamma_{m,e}^P.$ The estimated values of the magnetic and electric critical exponents are listed in Table \ref{table:I} and \ref{table:II}, respectively.

\subsection{Cluster properties}
A crucial part of studying percolation lies in understanding the properties of clusters. Although the density of the largest cluster characterizes the criticality at $T=1$, the other properties of the clusters also illustrate the criticality and scaling behavior. At critical temperatures, the distribution for cluster size shows power-law behavior.
\begin{figure}[t]
 \centering
 \includegraphics[width = 0.485 \linewidth]{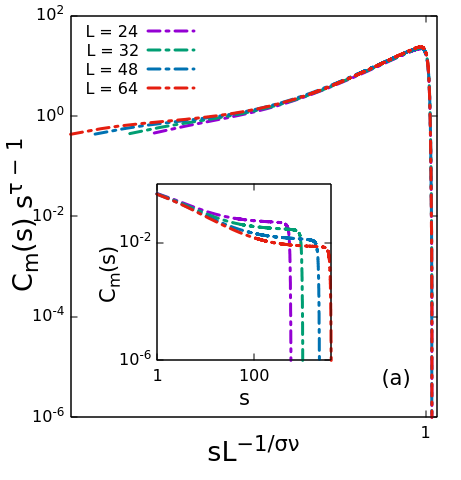}
 \includegraphics[width = 0.485 \linewidth]{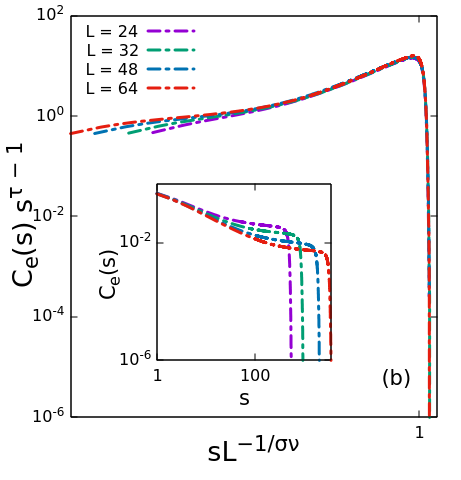} 
 \includegraphics[width = 0.485 \linewidth]{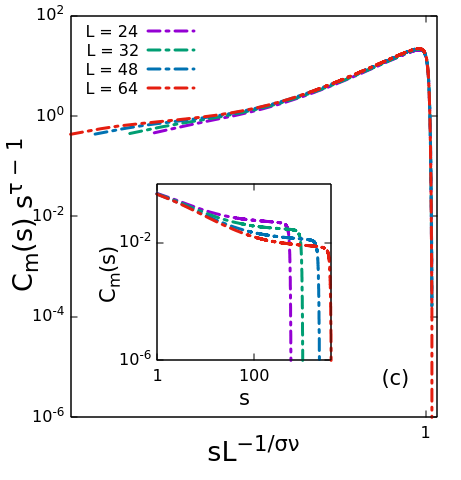} 
 \includegraphics[width = 0.485 \linewidth]{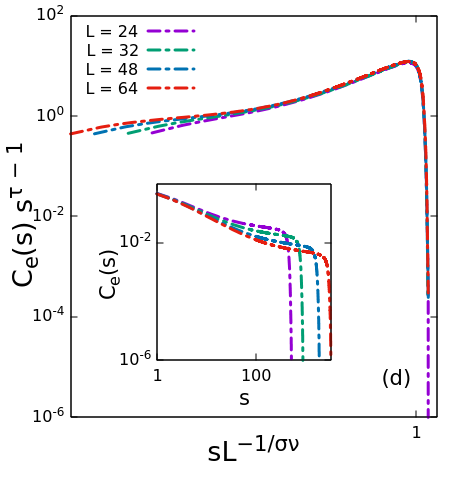} 
 \caption{ (Color online) Estimates of cluster exponents $\tau, \sigma$ from the scaling form of $C(s)_{m,e}$ in Eq. \eqref{cumulative-dist}. CDF of (a) magnetic clusters and (b) electric clusters for $\lambda = 0.1,$ results in best collapse for $\frac{1}{\sigma\nu} = 1.94(8),1.91(0)$ and $\tau= 2.0(3),2.0(4).$ (c),(d): the same for $\lambda = -0.1,$ with exponents $\frac{1}{\sigma\nu} = 1.94(8), 1.88(3)$ and $\tau= 2.0(3),2.0(6).$ Inset in all figures shows the raw data, averaged over $10^5$ samples in the steady state. Superscripts in $\tau, \sigma$ are dropped for better visibility. }  
 \label{fig:distribution}
\end{figure}
If $P_{m,e}(s)$ is the probability that a randomly chosen site belongs to a cluster of size $s$, at the critical point, it scales as $P_{m,e}(s) = s^{-\tau_{m,e}} f_{m,e}(\epsilon s^{\sigma})$. Consequently, using the scaling relation between $\epsilon$ and $L$, one can write $P_{m,e}(s) = s^{-\tau_{m,e}} K_{m,e} (sL^{-1/\nu\sigma_{m,e}})$. Thus, the cumulative density function (CDF) can be written as
\be
C_{m,e}(s) = \sum_{s' = s}^{\infty} P_{m,e}(s') = s^{1-\tau_{m,e}} K_{m,e}(sL^{-\frac{1}{\nu\sigma_{m,e}}}).\label{cumulative-dist}
\ee
The CDF of the magnetic and electric clusters are presented in Fig. \ref{fig:distribution} for $\lambda = \pm 0.1$. In these figures we plot $C(s)_{m,e} s^{\tau_{m,e}-1}$ as a function of $sL^{-\frac{1}{\nu\sigma_{m,e}}}$ and tune the exponents $\tau_{m,e}$
$(\nu\sigma_{m,e})^{-1}$ to get the best collapse. The resulting value of the exponents is given in the caption there.
These exponents are related to the static exponents $\beta$, $\gamma$ and $\nu$ through scaling relations
\cite{Stauffer_1992Book},
\bea
\tau_{m,e}&& = 2+\frac{\beta_{m,e}^P}{ \beta_{m,e}^P+ \gamma_{m,e}^P};~ \sigma_{m,e}^{-1} = %\frac{1}
{\beta_{m,e}^P + \gamma_{m,e}^P}.
%\cr&&~~~~~~~ D_{m,e} = d - \frac{\beta_{m,e}^P}{\nu}.
 \label{eq:scaling-II}
\eea
From the static exponents listed in Table \ref{table:I}, \ref{table:II}, we calculate $\tau, (\sigma\nu)^{-1}$ using the above scaling relations and verify for magnetic and electric percolation and verify that in all cases, for both magnetic and electric percolation, they agree well with the respective values estimated here independently.

%%%%%%%%%%%%%%%%%%%%%%%%%%%%%%%%%%%
\subsection{Percolation exponents on Baxter line}
%%%%%%%%%%%%%%%%%%%%%%%%%%%%%%%%%%%
%%%%%%%%%%%%%%%%%%%%%%%%%%%%%%%%%%%

\begin{figure}[t]
 \centering 
 \includegraphics[width = 0.485 \linewidth]{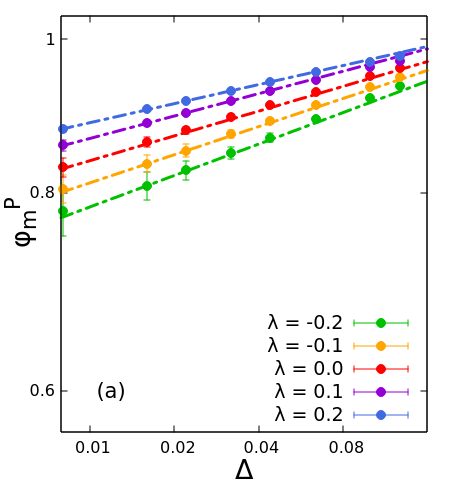}
 \includegraphics[width = 0.485 \linewidth]{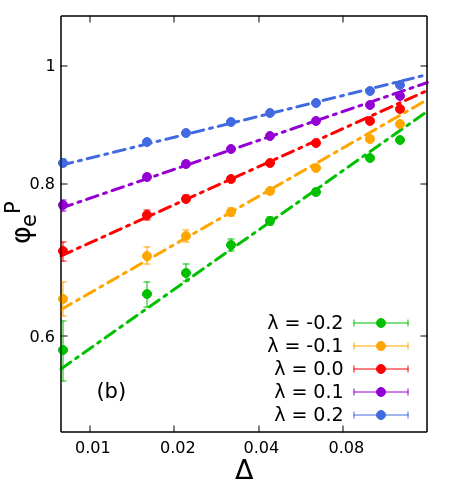}
 \includegraphics[width = 0.485 \linewidth]{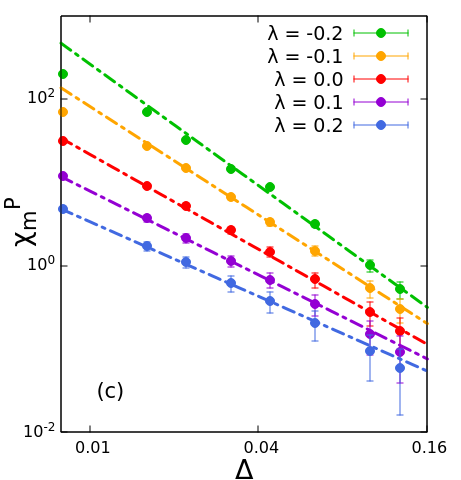} 
 \includegraphics[width = 0.485 \linewidth]{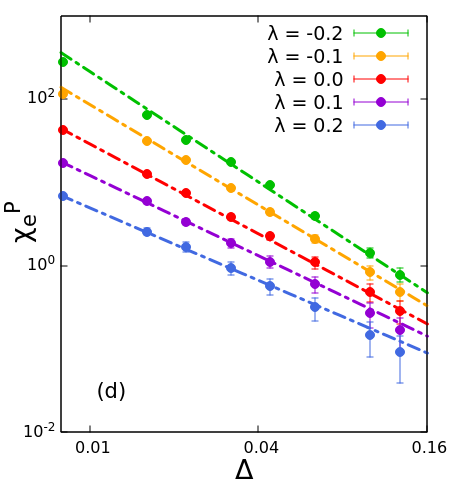} 
 \caption{ (Color online) For a large system $L = 512,$ critical exponents $\beta_{m,e}^P$, $\gamma_{m,e}^P$ are calculated from log scale plots of $\phi_{m,e}^P$ in (a),(b), and $\chi_{m,e}^P$ in (c),(d), as a function of $\Delta=T_c-T.$ $10^6$ steady state configurations are are considered.}
 \label{fig:direct-fit}
\end{figure}
%p'eExp.dat' u 1:(2-5*$3/$2/12) w lp,'eExp.dat' u 1:(2-$5/$2) w lp,'eExp.dat' u 1:(1+$6/$2/2) w lp

\begin{figure}[h]
 \centering 
 \includegraphics[width = 0.485 \linewidth]{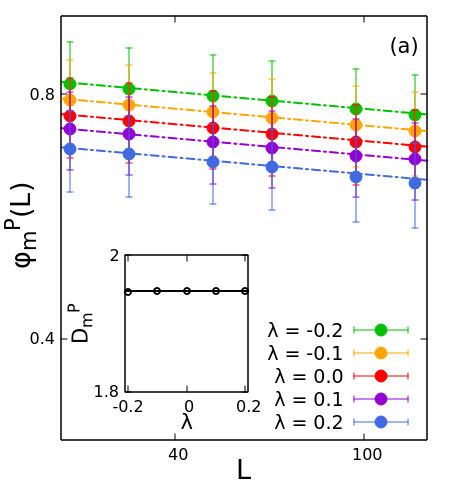} 
 \includegraphics[width = 0.485 \linewidth]{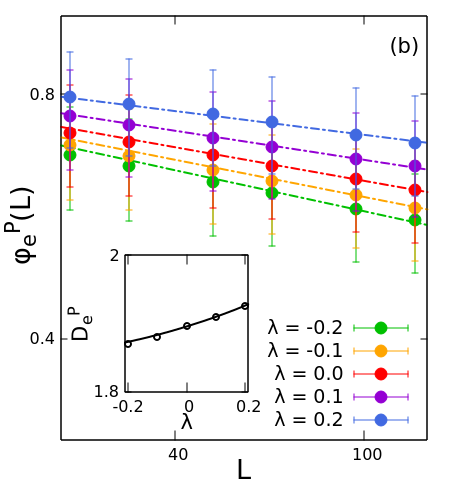} 
 \caption{(Color online) Estimation of fractal dimension $D_{m,e}=\log_{L}(s_{max}^{\tau, \alpha})$ of critical spanning clusters. Since, $s_{max}^{\tau, \alpha} = \phi_{m,e}L^2$, we plot $\phi_{m,e}$ vs. $L$ in log-scale to get the slopes $D_{m,e}.$ (a) Magnetic clusters: $D_m \approx 1.948$ for all $\lambda$ values (see inset) (b)Magnetic clusters: $D_e$ varies with $\lambda,$ shown in the inset. $10^7$ more clusters are considered in each case.}
 \label{fig:df}
\end{figure}

\begin{figure}[t]
 \centering
 \includegraphics[width = 0.485 \linewidth]{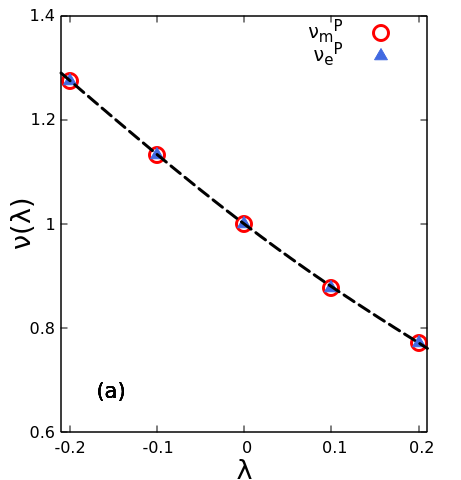}
 \includegraphics[width = 0.485 \linewidth]{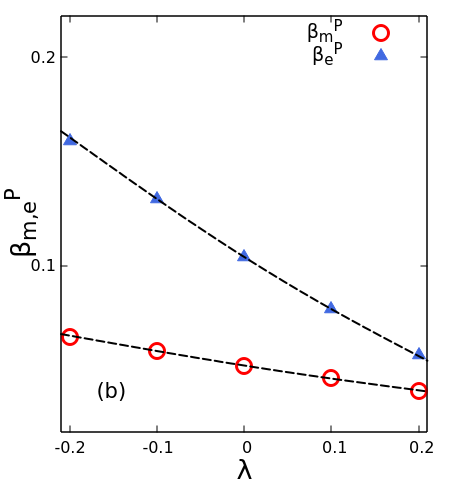}
 \includegraphics[width = 0.485 \linewidth]{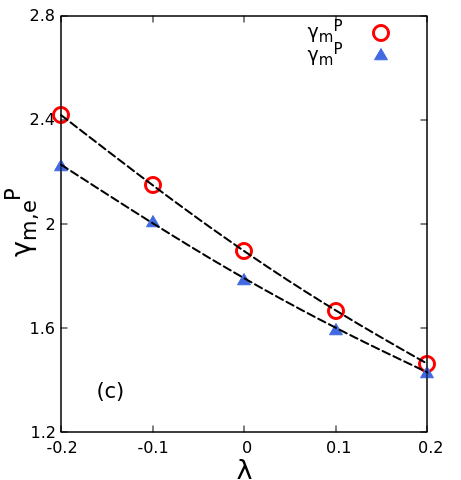}
 \includegraphics[width = 0.485 \linewidth]{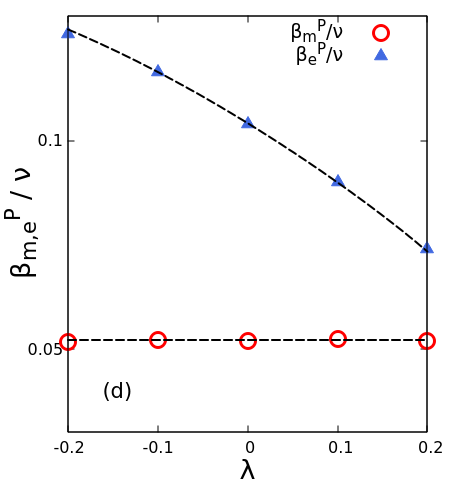}
 \caption{Critical exponents estimated from simulations (points) along with functions (line) predicted in Eq. \eqref{eq:exact_expP}. }
 \label{fig:expo}
\end{figure}

In the previous sections, we have reported on ordinary critical exponents $z,\nu,\beta,\gamma$ and the cluster exponents $\tau,\sigma$ for $\lambda=0.1.$ Exponents for other values of $\lambda$ are obtained in a similar way using finite-size scaling analysis. The respective figures for other $\lambda$ values are given in the Supplemental Material \cite{SM} to avoid repetition. Our final estimate of the critical exponents for $\lambda=-0.2, -0.1, 0, 0.1, 0.2$ are listed in Table \ref{table:I} and Table-\ref{table:II}, for magnetics and electric percolation respectively.

To ensure that the finite size corrections are minimal in the FSS analysis, we measure $\beta_{m,e}^P$ and $\gamma_{m,e}^P$ for a large system $L=512,$ from the
log-scale plots of $\phi^P_{m,e}, \chi^P_{m,e}$ as a function of $\Delta.$
This is shown in
Fig. \ref{fig:direct-fit} for different $\lambda$ values - the straight line with slope $\beta_m,\beta_e,\gamma_{m},\gamma_e$ listed in Tables \ref{table:I} \ref{table:II} are drawn along with the data in Figs.\ref{fig:direct-fit} (a), (b), (c), (d) respectively to guide the eye. A very good match makes us confident about our estimates of critical exponents from FSS.

{\it Exact values of the exponents:} To know the exact values of the critical exponents we rely on an old conjecture \cite{Stella_1989} that fractal dimension $D^P$of the geometric clusters of Ising model can not be determined solely in terms of the Ising exponents \eqref{eq:Ising}, there is an additional factor $w=\frac{5}{12}$ which comes from the connection of Ising percolation with tricritical $q=1$ Potts model \cite{Stella_1989}. We conjecture that the same holds for magnetic and electric percolation,
\be
D_{m,e} = d - w \frac{\beta_{m,e}}{\nu}; w= \frac{5}{12}
\label{eq:D1}
\ee
where $d=2$ and $\beta_{m,e}$ are given in Eq. \eqref{eq:exact_exp}. The fractal dimension is defined as how the largest cluster at criticality
scales with the system size as
 \be \bra{s_{max}^{\tau,\alpha}} \sim L^{D_{m,e}} \implies \phi_{m,e}=\frac{\bra{s_{max}^{\tau,\alpha}}}{L^d} \sim L^{D_{m,e} -d}.
 \label{eq:Dd}
 \ee
But for a finite system at the critical point $ \phi_{m,e}$ scales as $L^{-\beta_{m,e}/\nu}.$ This results in a relation,
 \be D_{m,e} = d - \frac{\beta_{m,e}^P}{\nu}.\label{eq:D2}\ee
Since the correlation length is an intrinsic property of the system, the associated exponent $\nu$ is invariant irrespective of whether we study magnetization, polarization, or their percolation properties. Thus, from Eqs. \eqref{eq:D1} and \eqref{eq:D2} we get $\beta_{m,e}^P= \frac{5}{12}\beta_{m,e}$ and $\gamma_{m,e}^P = d\nu -\beta_{m,e}^P.$ Explicitly, the percolation exponents vary with $\lambda$ as,
\bea
&\nu_{m,e}^P =\nu= \frac{2(\mu - \pi)}{(4\mu - 3\pi)} \text{ where } \cos\mu = e^{2\lambda}\sinh{(2\lambda)};\cr
&\beta_m^P = \frac{5\nu}{96};~\beta_e^P = \frac{5(2\nu-1)}{48}; \gamma_m^P = \frac{91\nu}{48}; \gamma_e^P= \frac{5 + 38 \nu}{24}.
\label{eq:exact_expP}
\eea

These results are based on the conjecture that the fractal dimension of spanning clusters at criticality is given by Eq. \eqref{eq:D1} and $\nu$ remains invariant. In Fig. \ref{fig:df} (a), (b) we plot $\phi_m$ and $\phi_m$ as a function of $L,$ for different $\lambda$ values on the Baxter line at $T=1;$ the slopes give us $D_{m,e}-2$ ( using $d=2$ in Eq. \eqref{eq:Dd}). The resulting value of $D_{m,e}$ are listed in
Tables \ref{table:I}, \ref{table:II} respectively. Note that, within the error limits, $D_m$ is the same for all $\lambda$, whereas $D_e$ varies. This is expected from Eq. \eqref{eq:D1}: if $w$ is independent of $\lambda,$ then for magnetic transition which obey weak universality with $\beta_m/\nu =1/8$ for all $\lambda,$ $D_m$ must be a constant $\frac{187}{96}\simeq 1.94792,$ which is consistent with our estimates. On the other hand, $\beta_e/\nu$ depends on $\lambda,$ so does $D_e.$ 

Next, we check if $\nu$ obtained from FSS of the Binder cumulants of $s_{max}^{\tau,\alpha}$ (and listed in Tables \ref{table:I}, \ref{table:II}) obey the functional form given in Eq. \eqref{eq:exact_exp}. This is shown in Fig. \ref{fig:expo}(a), where two different symbols are our estimates from electric and magnetic percolation, and the line corresponds to the exact expression. A good match indicates that $\nu$ is indeed invariant.

The theoretical values of $\beta_{m,e}^P, \gamma_{m,e}^P$ obtained based on this conjecture in Eq. \eqref{eq:exact_expP} are compared
with the respective values obtained in Tables \ref{table:I}, \ref{table:II}), Figs. \ref{fig:expo}(b),(c); we find that they agree well. Finally in Fig. \ref{fig:expo}(d) we compare $\frac{\beta_{m,e}}\nu= d- D_{m,e}$ with the respective theoretical values.

These results indicate very strongly that the percolation exponents on the Baxter line are given by Eq. \eqref{eq:exact_expP}. The crucial point is that, although all the critical exponents vary with $\lambda,$ the fractal dimension of the spanning cluster at criticality is related to the respective underlying magnetic and electric phase transition via a universal constant $w = \frac{5}{12}$ this number $\beta_m^P/ \beta_m =w= \beta_e^P/ \beta_e$ is unique constant for a given universality class. Recently, it was reported that there are superuniversal functions \cite{Indranil_PRB2023} at the critical point, which remains invariant along any marginal directions where exponents vary continuously. We discuss the superuniversality class of the percolation transition along the Baxter line in section IV.

%%%%%%%%%%%%%%%%%%%%%%%%%%%%%%%%%%%%%%%%
\section{Percolation at $Z_4$ }
%%%%%%%%%%%%%%%%%%%%%%%%%%%%%%%%%%%%%%%%%%%%%
%The point $\lambda = \frac{1}{4}\ln(3)$ on Baxter line is special as there we have $J=\lambda.$ 
The isotropic AT model is special when $J=\lambda;$ here the Hamiltonian is invariant under the permutation of four states $(\{\sigma = \pm\},\{\tau = \pm\}).$ On the Baxter line this symmetry is observed when $\lambda = \frac{1}{4}\ln(3)=J$ and there the critical behavior is equivalent to that of $Z_4$ symmetry breaking concurring in the $4$-state Potts model \cite{Wu_Potts_RMP_1982}. This point stands out of the entire Baxter line as the electric and magnetic transitions are no longer distinguishable respective critical exponents ( listed in Eq. \eqref{eq:Z4exp}) are identically the same. The corresponding percolation transition must belong to the percolation transition in the $4$-state Potts model.

To calculate cluster properties at the $Z_4$ point, we define
$S_{\bf i} = \frac{3+ \sigma_{\bf i}}2 + \tau_i$ at site every site ${\bf i}$ so that all four equivalent states ${(\sigma_{\mathbf{i}},\tau_{\mathbf{i}}) = (\pm, \pm)}$ are denoted by different integers $S_{\bf i}=0,1,2,3.$ The clusters are formed by connecting nearest neighbors which have the same $S$- values. Thus, if a configuration has $K$ clusters, labeled by $k=1,2\dots,K$ 
and size or mass of $k$-th cluster is $s_k$ we have $\sum_{k=1}^K s_k= L^2.$ In every configuration, the size of the largest cluster is denoted by $s_{max}.$ The order parameter for percolation transition at $Z_4$ point is then, $\phi_4^P = \frac{1}{L^2}\langle s_{max} \rangle.$
Then, the static critical exponents are defined by the following relations.
\be
\phi_4^P\sim \Delta^{\beta_4^P};~\chi_4^P = \frac{\langle {s_{max}}^2\rangle - \langle s_{max}\rangle^2}{L^4} \sim \Delta^{\gamma_4^P} 
\ee
Percolation in the 4-state Potts model has been previously studied analytically \cite{Janke_1995}. The fractal dimension $D_4$ of the geometric clusters at the critical point is related to the ordinary
critical exponents as $D_4 = d- w \frac{\beta_4}{\nu_4}$ with $w=1.$
From $w=1$ it it is clear that the percolation exponents at $Z_4$ point must be the same as that of the ordinary critical exponents for
$Z_4$-symmetry breaking.
\be 
\nu = \frac{2}{3};~\beta_4^P = \frac{1}{12};~\gamma_4^P = \frac{7}{6}
\label{eq:Z4Pexp}
\ee

\begin{figure}[h]
 \centering
 \includegraphics[width = 0.485 \linewidth]{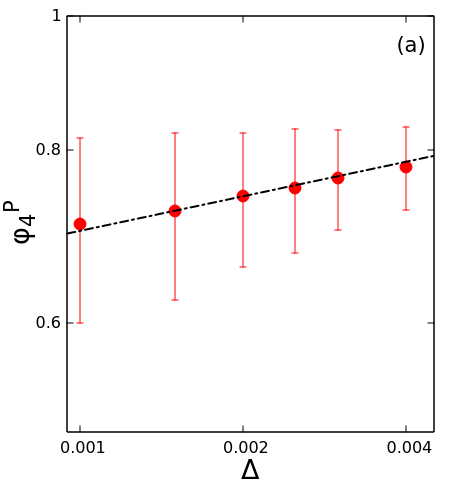}
 \includegraphics[width = 0.485 \linewidth]{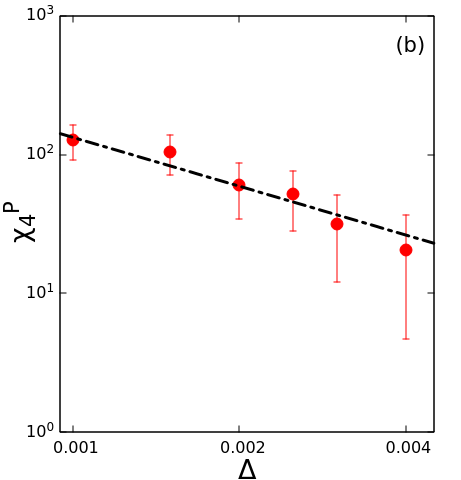}
  \includegraphics[width = 0.485 \linewidth]{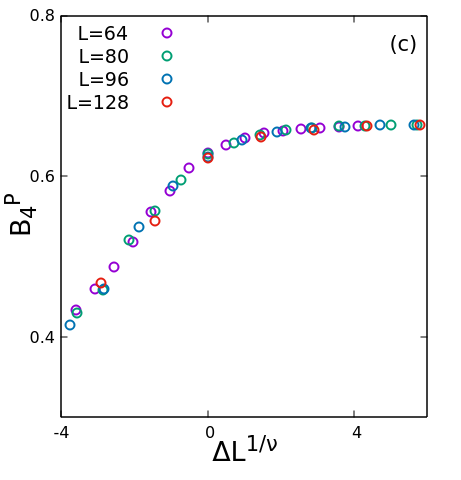}
 \includegraphics[width = 0.485 \linewidth]{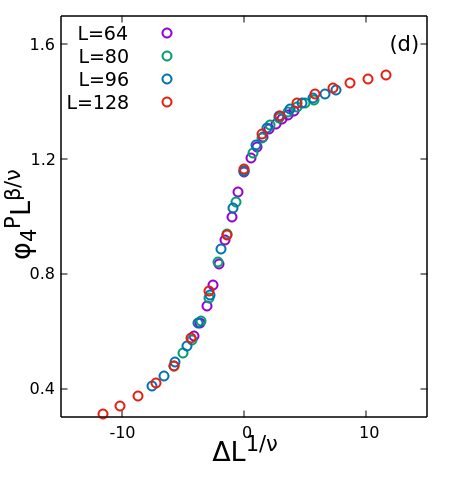}
 \caption{ (Color online) Critical percolation at $Z_4$ ($\lambda = \frac{1}{4}\ln 3$). Log-scale plot of (a)$\phi_4$ and (b)$\chi_4$ as a function of $\Delta=T_c-T$ (system size $L=128$). Lines with slopes $\beta_4^P = \frac{1}{12}$ and $\gamma_4^P = \frac{7}{6}$ are drawn along the data (points) for comparison. Estimate of (c) $\frac1\nu$ from data collapse of Binder cumulant $B_4^P$ vs. $\Delta L^{1/\nu}$ and (d) $\frac\beta\nu$ from collapse of $L^{\beta/\nu} \phi_4^P$ vs. $\Delta L^{1/\nu}$ for different $L.$ The relevant variables are averaged over $10^7$ samples.}
 \label{fig:q4}
\end{figure}

At $\lambda = \frac{1}{4}\ln(3)$, the finite-size correction to the scaling is extremely large. This affects the estimation of the critical exponents strongly, in FSS analysis. In Fig. \ref{fig:q4} (a) and (b) we show a log-scale plot of the percolation order parameter $\phi_4^P$ and susceptibility $\chi_4^P$ obtained from Monte Carlo simulations as a function of $\Delta= T_c-T.$ Straight lines of slope $\beta_4^P = \frac{1}{12}$ and $\gamma_4^P = \frac{7}{6}$ are drawn along with the data. The error bars are very large here because of the strong finite size effect; but the data is consistent with predictions in Eq. \eqref{eq:Z4Pexp}.

\section {Superuniversality} 

\begin{figure}[h]
 \centering 
 \includegraphics[width = 0.985 \linewidth]{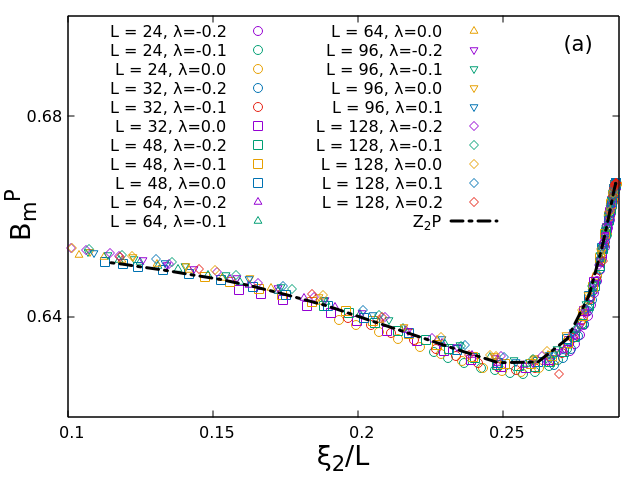} 
 \includegraphics[width = 0.985 \linewidth]{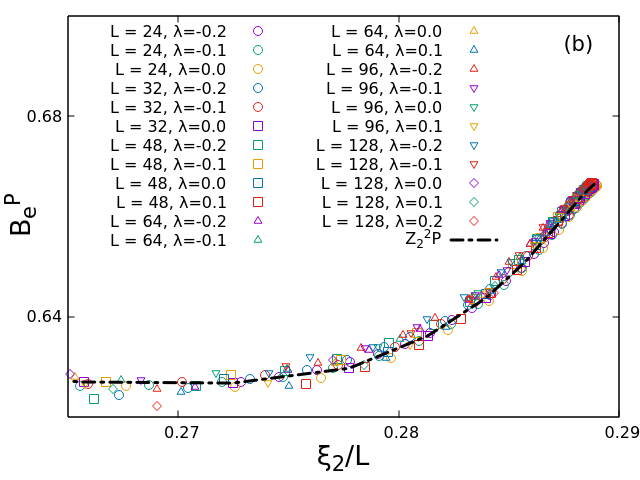}
\caption{(Color online) All along the Baxter line, Binder cumulant is a unique function of $\frac{\xi_2}{L}.$ For magnetic percolation this function $B_{m}^P(\frac{\xi_2}{L})$ is same as that of the Ising percolation ($Z_2$P universality). $B_{e}^P(\frac{\xi_2}{L})$ in electric percolation matches with that of the overlapping Ising clusters($Z_2^2$P universality).}
 \label{fig:suh_xi2}
\end{figure}
\begin{figure}[h]
 \centering
 \includegraphics[width = 0.985 \linewidth]{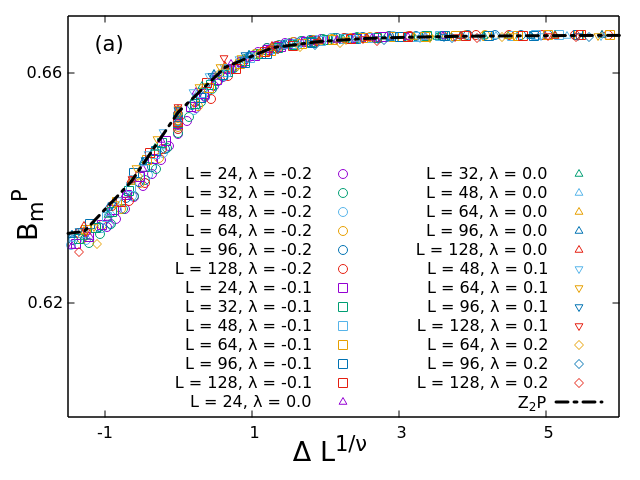} 
 \includegraphics[width = 0.985 \linewidth]{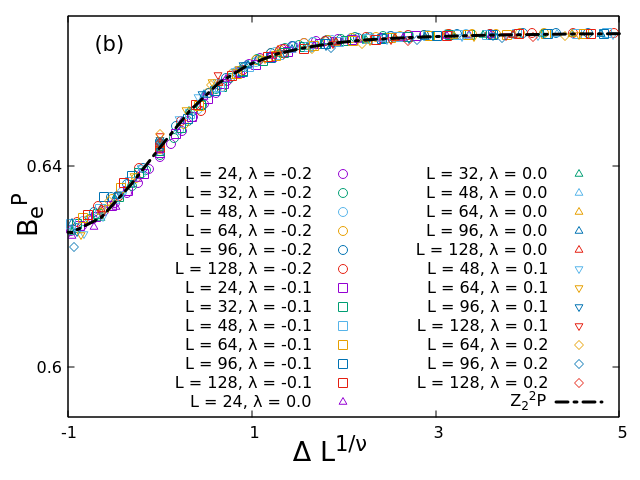} 
 \caption{(Color online) Collapse of scaling functions $B_{m,e}^P(\Delta L^{\frac{1}{\nu(\lambda)}})$. We average the necessary parameters over $10^7$ samples in the steady state.}
 \label{fig:suh_xi1}
\end{figure}

In the AT-model, both magnetic and electric percolation transitions show continuous variation of critical exponents along the Baxter-line (as shown in Tables. \ref{table:I}, \ref{table:II} respectively) and they obey scaling relations \eqref{eq:scaling-I} and \eqref{eq:scaling-II}. These features are identical to the underlying magnetization and Polarization transitions. But it has been shown recently \cite{Indranil_PRB2023} that the magnetic and electric transitions in AT model exhibit superuniversality in the sense that while the critical exponents vary according to a functional form that adheres to generic scaling relations, some of the scaling functions themselves remain invariant along the Baxter line.
Since at $\lambda = 0,$ the AT-model reduces to two independent Ising models, invariant scaling functions are identical to $Z_2$ and
$Z_2^2$ universality. It remains to see, whether percolation transitions also satisfy the superuniversality hypothesis.

To verify the superuniversality hypothesis we consider the RG-invariant scaling functions,
\be
B_{m,e}^P = {\cal G}_{m,e}(\frac{\xi_2}L)
\ee
where ${\cal G}_{m,e}$ are the super universal scaling functions that do not depend on the value of the critical exponents, the critical threshold values $J_c,\lambda_c$, or the system size. Here, $B_{m,e}^P$ are the Binder cumulants and $\xi_2$ is the second-moment correlation length defined in Eq. \eqref{eq:xi2}.

In Fig. \ref{fig:suh_xi2} (a) we plot $B_{m}^P$ as a function of $\xi_2/L$ for different values of $\lambda$ (and taking corresponding $J$ on the Baxter line) and different system size. All the data fall on a super-universal curve confirming that the magnetic percolation belongs $Z_2P$ super universality class.
A similar plot of $B_{e}^P$ as a function of $\xi_2/L$ is shown in Fig. \ref{fig:suh_xi2}(b). Data collapse is observed in a large range of parameters, again confirming that the electric-percolation transition forms a new universality class, namely $Z_2^2P.$ 

In addition, we investigate the behavior of the Binder cumulants $B_{m,e}^P$ as a function of $\xi/L$ which is equivalent to studying $B_{m,e}^P$ as a function of $ (\frac{\xi}L)^{-1/\nu} \equiv \Delta L^{1/\nu}.$ 
This is shown in Figs. \ref{fig:suh_xi1} (a), (b) respectively for magnetic and electric transition for different $\lambda, L.$ Here too, data for all parameters fall on a super-universal scaling function
%${\cal Q}_{m,e},$ 
\be
B_{m,e}^P = {\cal Q}_{m,e}(\Delta L^{1/\nu}).
\ee

\section{Conclusion}

In this article, we study the geometric percolation of spins in the Ashkin-Teller model, which is a bi-layer spin model with Ising spins $\tau$ and $\sigma$ interacting ferromagnetically in each layer (interaction strength $J$) with four-spin inter-layer interaction (strength $\lambda$) as in Eq. \eqref{hamiltonian}. We find that the largest of the geometric clusters of $\tau$ or $\sigma$ spins, and that of the spin-dipoles $\alpha=\tau\sigma,$ which is a spin-like variable at each site, spans the entire system simultaneously when interaction parameters $\lambda, J$ crosses a critical threshold. The critical line on the $\lambda$-$J$ plane matches with the Baxter line, which describes a transition of the system from being an ordinary paramagnet to a polarized ferromagnet. The magnetization and polarization transition
characterized by order parameters $\bra{\tau}= \bra{\sigma}$ and $\bra{\alpha}$ respectively.

\begin{figure}[h]
 \centering
 \includegraphics[width=7cm, height=6.5cm]{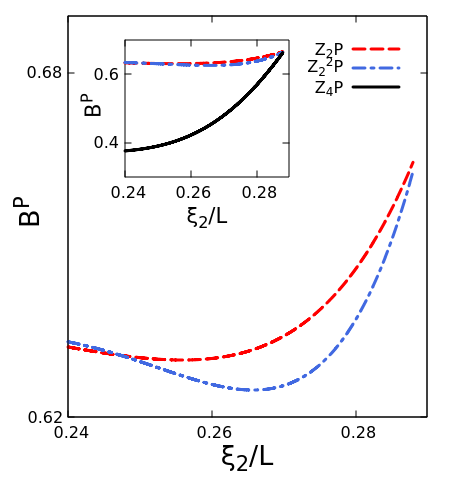}
 \caption{(Color online) Comparison of the superuniversal functions --Binder cumulant as a function of $\frac{\xi_2}{L}$-- of $Z_2$P and $Z_2^2$P universality class. The inset compares them with that of $Z_4$P.}
 \label{fig:compare_suh}
\end{figure}

The magnetic clusters are constructed separately in each layer by joining neighbors having parallel spins. In the electric clusters, neighboring sites are connected when they have the same $\alpha$ values. In our study, we follow the standard prescriptions of site-percolation transition to define the order parameters $\phi_{m,e}$ as the average size of the largest clusters $ \phi_{m,e}= \bra{s_{max}^{\tau, \alpha}},$ where subscripts $m,e$ stands for magnetic and electric percolation. Since $ \bra{s_{max}^\tau} = \bra{s_{max}^\sigma}$ we consider only $\tau$ spins for characterizing magnetic percolation transition. The critical exponents of both transitions are then obtained at different points on the Baxter line using finite-size scaling analysis. We calculate both static and dynamic exponents and in some cases, the exponents $\tau_{m,e}$ and $\sigma_{m,e}$ related to the cluster size distribution; they are listed in Tables \ref{table:I} and \ref{table:II}.

We find that the critical exponents $\beta_{m,e}, \gamma_{m,e}, \nu$ depends on $\lambda$ the Baxter line also obey the known scaling relations \eqref{eq:scaling-I}; the cluster exponents too obey scaling relations \eqref{eq:scaling-II}. A plot of the exponents as a function of $\lambda$ is shown in Fig. \ref{fig:expo} along with exact values obtained in Eq. \eqref{eq:exact_expP} following the conjecture \eqref{eq:D1}; they agree remarkably well and justify the validity of the conjecture.

Finally, we ask if the transitions belong to the superuniversality class of correlated percolation or Ising percolation ($Z_2$P). It was shown in a recent work that Binder cumulant as a function of $\xi_2/L$ with $\xi_2$ being the second moment correlation length is a superunivarsal function ${\cal G}( \frac{\xi_2}L)$ that remain invariant all along a critical line formed by a marginal operator; this invariance is crucial in determining the universality class as the set of critical exponents vary along the critical line. In this study we find that magnetic and electric percolation too obey this super universal hypothesis with scaling functions ${\cal G}_{m,e}(.)$; ${\cal G}_{m}(.)$ is same as that of Ising percolation universality class $Z_2$P known earlier and ${\cal G}_{e}(.)$ form a new superuniversality class, $Z_2^2$P that corresponds to percolation of spin-dipoles or electric percolation. A recent study on soft geometric clusters of overlapping Ising model studied in the context of spin glass transition, has obtained the same critical exponents (within error limits) we report here for electric percolation at $\lambda=0;$ then, percolation of overlapping Ising clusters belongs to the $Z_2^2$P universality class.

Further $\lambda= \frac14 \ln(3)$ is a special point on the Baxter line which has $Z_4$ symmetry. Here, percolation of geometric clusters belongs to the $Z_4$ universality class associated with the $4$-state Potts model. The superuniversal function ${\cal G}_{4}(.)$ at $Z_4$ point is different from both ${\cal G}_{m}(.)$ and ${\cal G}_{e}(.);$ All three super universal functions are compared in Fig. \ref{fig:compare_suh}.

\begin{table}[t]
\caption{ Critical exponents of magnetic percolation \label{table:I}}
\begin{tabular}{|c|c|c|c|c|c|c|c|c|} 
 \hline
$\lambda$ & $z$ & $\nu_m^P$ &$\beta_m^P$ &$\gamma_m^P$ & $\tau_m^P$&$\sigma_m^P$&$w$\footnote{$w=\beta_m^P/\beta_m$; $\beta_m$ taken from Eq. \eqref{eq:exact_exp}.} &$D_m$\\ \hline
-0.2& 2.0(0) &1.27(5) &0.06(6) &2.4(2) & 2.0(3) &0.40(3) &0.415 & 1.947\\ 
-0.1& 2.0(0) &1.13(3) &0.05(9) &2.14(3) & 2.0(3) & 0.45(3) &0.415 & 1.948\\ 
0.0& 2.0(0) &1.00(0) &0.052(1)&1.89(6) & 2.0(3) & 0.51(3) &0.416 & 1.948\\ 
0.1& 2.0(0) &0.87(8) &0.04(6) &1.66(3) & 2.0(3) & 0.58(5) &0.422 & 1.948\\ 
0.2& 2.0(1) &0.77(1) &0.04(0) &1.46(2) & 2.0(3) & 0.66(6) &0.417 &1.948\\ 
\hline
\end{tabular}
\end{table}
%%%%%%%%%%%%%%%%%%%%%%%%%%%%%%%%%%%
\begin{table}[t]\caption{ Critical exponents of electric percolation \label{table:II}}
\begin{tabular}{|c|c|c|c|c|c|c|c|c|} 
 \hline
$\lambda$ & $z$ & $\nu_e^P$ &$\beta_e^P$ &$\gamma_e^P$ & $\tau_e^P$& $\sigma_e^P$& $w$\footnote{$w=\beta_e^P/\beta_e$; $\beta_e$ taken from Eq. \eqref{eq:exact_exp}.} & $D_e$\\ \hline
-0.2& 2.0(0) &1.27(5) &0.16(0) &2.22(0) &2.0(8) &0.41(9) &0.412 &1.87\\
-0.1& 2.0(0) &1.13(3) &0.13(2) &2.00(2) &2.0(6) &0.46(8) &0.418 &1.88\\ 
0.0& 2.0(0) &1.00(0) &0.10(4) &1.78(0) &2.0(5) &0.52(7) &0.416 &1.90\\ 
0.1& 2.0(5) &0.87(8) &0.07(9) &1.58(8) &2.0(5) &0.59(6) &0.418 &1.91\\
0.2& 2.1(6) &0.77(1) &0.05(7) &1.42(6) &2.0(4) &0.67(3) &0.422 &1.926\\ 
\hline
\end{tabular}
\end{table}

A summary of our findings follows. From Monte Carlo simulations we find critical exponents of percolation of spins and spin dipoles occurring along the Baxter line in the Ashkin-Teller model and show that their functional form matches with the exact values conjectured in \eqref{eq:exact_expP}. Critical exponents of magnetic percolation (percolation of spins in individual layers) vary continuously following weak universality (some exponent ratios are fixed) whereas the same for electric percolation (percolation of spin dipoles) are fully non-universal. All along the line, the order parameter exponents of both percolation are a constant multiple of respective exponents of magnetization and polarization; this constant $w$ is a universal constant associated with a universality class, here $w=\frac5{12}$ associated with Ising universality. That $\beta$ for percolation is a constant multiple of the order parameter exponent of underlying transition is quite useful in determining the universality class of a phase transition from studying the percolation transition there. Recently this idea has been used in the study of motility-induced phase transition of run and tumble particles on a lattice \cite{Kanti_2024}.

We also find that the super universal function, Binder cumulant as a function of $\xi_2/L$ remains invariant along the Baxter line for both percolation and matches with the same function obtained for $Z_2$P (Ising percolation) and $Z_2^2$P (percolation of spin dipoles) respectively indicating that the magnetic, electric percolation along the Baxter line belong respectively to $Z_2$P and $Z_2^2$P superuniversality class.

\begin{acknowledgments}
The authors like to acknowledge helpful discussions with Indranil Mukherjee and Sayantan Mitra.
\end{acknowledgments}

\bibliography{perco}
\bibliographystyle{apsrev4-2}

\end{document}

\title{Supplemental Material: Geometric percolation  of spins and spin-dipoles in Ashkin Teller model}

\author{ 
Aikya Banerjee, Priyajit Jana and P. K. Mohanty}
\email{pkmohanty@iiserkol.ac.in}
\affiliation {Department of Physical Sciences, Indian Institute of Science Education and Research Kolkata, Mohanpur, 741246 India.}

\begin{abstract}
In  this Supplemental Material, we report on estimation
of percolation critical exponents for different points on
Baxter line in Ashkin-Teller model.
\end{abstract}
\maketitle

Ashkin Teller model  is a two layer lattice  model in   with Ising spins  $\sigma_{\bf i}$ and $\tau_{\bf i}$ at  site ${\bf i}\equiv (x,y),$
$x,y= 1,2,\dots, L$  interacting  following  the  Hamiltonian, 
\begin{equation} 
\mathcal{H}= - J\sum_{\bra{{\bf i},{\bf j}}} \left( \sigma_{\bf i} \sigma_{\bf j} +\tau_{\bf i} \tau_{\bf j}\right) - \lambda \sum_{\bra{{\bf i},{\bf j}}} \sigma_{\bf i}\tau_{\bf i} \sigma_{\bf j}\tau_{\bf j}.
\label{hamiltonian}
\end{equation}
The interaction  term can be interpreted as the   Ising interaction of strength $\lambda$ among spin dipoles $\alpha_{\bf i} = \sigma_{\bf i}\tau_{\bf i}.$ At  temperature $T=1,$ the model  exhibits a continuous phase transition  along a   critical line 
\be
J_c  = \frac12\sinh^{-1}(e^{-2 \lambda}) \label{eq:Jc}
\ee
in $\lambda$-$J$ plane when $\lambda<\lambda^*= \ln(3)/4.$  The critical exponents of magnetization $M= \bra{\sigma_{\bf i}}=\bra{\tau_{\bf i}}$ and polarization $P=\bra{\alpha_{\bf i}}$ occurs together on the Baxter line with critical exponents, known exactly from the mapping of the model to the eight vertex model,  
\bea
&\nu = \frac{2(\mu - \pi)}{(4\mu - 3\pi)}, \text{ where } \cos\mu = e^{2\lambda}\sinh{(2\lambda)};~
\beta_m = \frac{\nu}{8}, \gamma_m = \frac{7\nu}{4}; \quad \beta_e = \frac{2\nu-1}{4}, \gamma_e = \frac{1}{2} + \nu,
\label{eq:exact_exp}
\eea
where subscripts $m,e$ stands for magnetic and electric  transitions.  

We aim at investigating percolation of spins (magnetic percolation) and spin dipoles (electric percolation) along the  Baxter line  to  determine the critical exponents and the universality class.  We conjecture that the fractal dimension of percolation clusters at criticality  are given  a relation,   $D_{m,e}= d- w \frac{\beta_{m,e}}{\gamma_{m,e}}$ with $w= \frac{5}{12},$   similar to  the one observed in percolation of spins in Ising model. This  results in continuous  variation of  percolation critical exponents,
\bea
&&\nu_{m,e}= \nu = \frac{2(\mu - \pi)}{(4\mu - 3\pi)}, \text{ where } \cos\mu = e^{2\lambda}\sinh{(2\lambda)};\cr
&&\beta_m^P = \frac{5\nu}{96};~\beta_e^P = \frac{5(2\nu-1)}{48};
\gamma_m^P = \frac{91\nu}{48}; \gamma_e^P = \frac{5 + 38 \nu}{24};~ 
\tau_{m,e} = 2+\frac{\beta_{m,e}^P}{  \beta_{m,e}^P+ \gamma_{m,e}^P};~ \sigma_{m,e}^{-1} = %\frac{1}
{\beta_{m,e}^P + \gamma_{m,e}^P}.
%\cr&&~~~~~~~ D_{m,e} = d - \frac{\beta_{m,e}^P}{\nu}.
\label{eq:exact_expP}
\eea
Which satisfy the  scaling relations, 
$D_{m,e} = d - \frac{\beta_{m,e}^P}{\nu};~   2\beta_{m,e}^P + \gamma_{m,e}^P.$

In the main text we verified the conjecture  and the  critical exponents  from Monte Carlo simulation of the model, mostly for the critical point  for $\lambda=0.1$ on Baxter line,  using  the finite size scaling properties of  the 
the order parameter, susceptibility and the binder cumulants, 
\be \phi_{m,e}^P \sim ; \frac{ \bra{s_{max}^{\tau,\alpha}}}{L^2};~
 \chi^P_{m,e} = \frac{ \bra{(s_{max}^{\tau,\alpha})^2} - \bra{s_{max}^{\tau,\alpha}}^2}{L^4}; ~ B_{m,e}^P = 1-   \frac{ \bra{(s_{max}^{\tau,\alpha})^4}}{ 3\bra{(s_{max}^{\tau,\alpha})^2}^2}
 \ee
 Here  we provide  data  and data collapse for other  $\lambda$ values,  including  some more results about  $\lambda=0.1.$ 
 The details  are  presented in the caption of each figure.

\begin{figure}
  
    \begin{subfigure}{0.245\linewidth}
    \includegraphics[width = \linewidth]{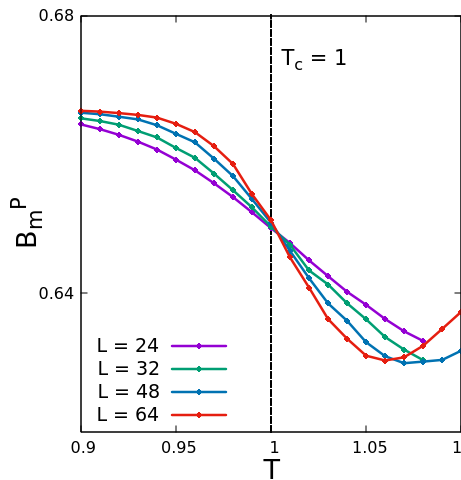}
    \subcaption{}
    \end{subfigure}
    \begin{subfigure}{0.245\linewidth}
    \includegraphics[width = \linewidth]{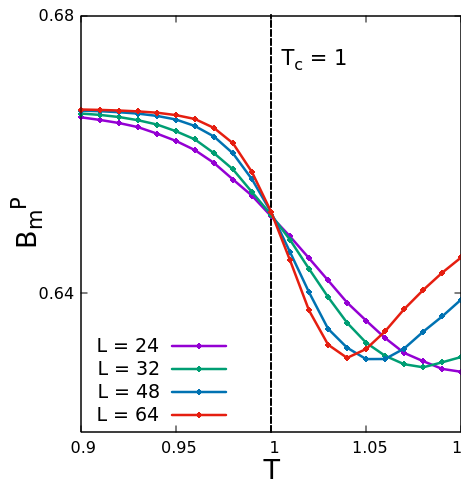}
    \subcaption{}
    \end{subfigure}
    \begin{subfigure}{0.245\linewidth}
    \includegraphics[width = \linewidth]{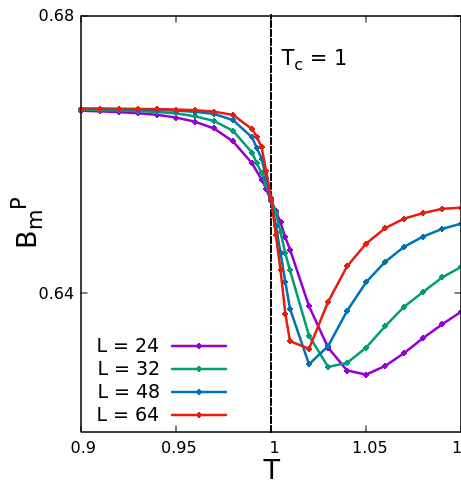}
    \subcaption{}
    \end{subfigure}
    \begin{subfigure}{0.245\linewidth}
    \includegraphics[width = \linewidth]{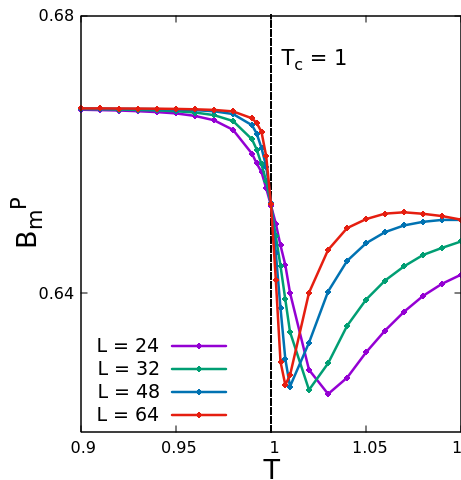}
    \subcaption{}
    \end{subfigure}
\captionsetup{justification=raggedright, singlelinecheck=false}
        \caption{Binder cumulant $B_m^P$  as a function of $T$  for different $L$   intersects at $T_c=1.$  (a) $\lambda = -0.2$, (b) $\lambda = -0.1$, (c) $\lambda = 0.0$, (d) $\lambda = 0.2.$  In all cases $J=J_c,$  given by Eq. \eqref{eq:Jc} and data are averaged over $10^7$ samples.}
    \label{fig:supp1}
\end{figure}

\begin{figure}
    \begin{subfigure}{0.245\linewidth}
    \includegraphics[width = \linewidth]{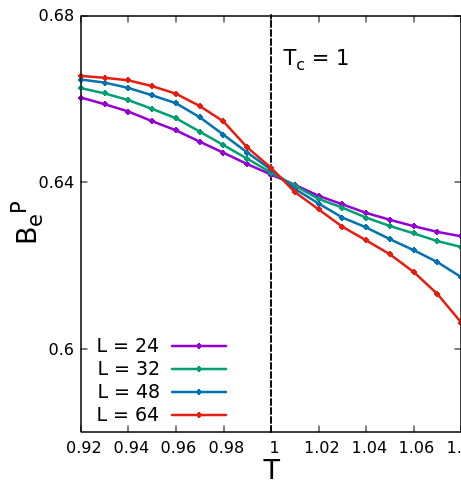}
    \subcaption{}
    \end{subfigure}
    \begin{subfigure}{0.245\linewidth}
    \includegraphics[width = \linewidth]{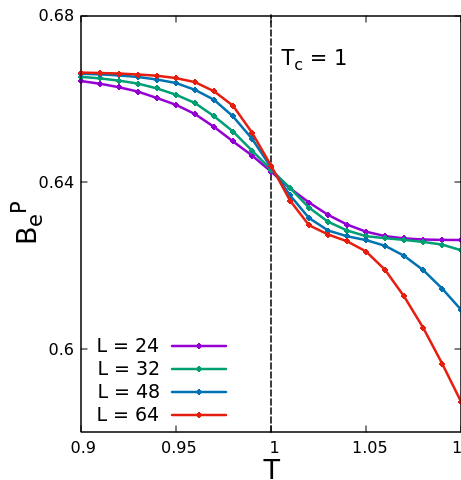}
    \subcaption{}
    \end{subfigure}
    \begin{subfigure}{0.245\linewidth}
    \includegraphics[width = \linewidth]{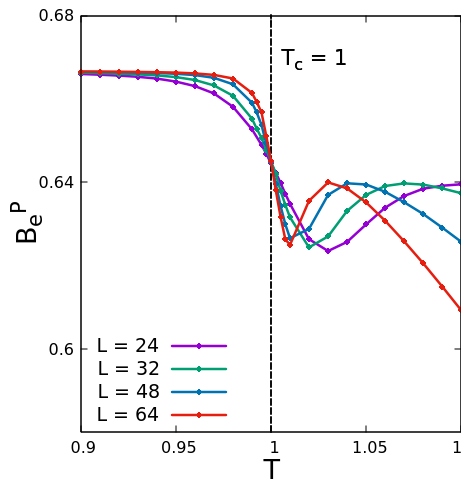}
    \subcaption{}
    \end{subfigure}
    \begin{subfigure}{0.245\linewidth}
    \includegraphics[width = \linewidth]{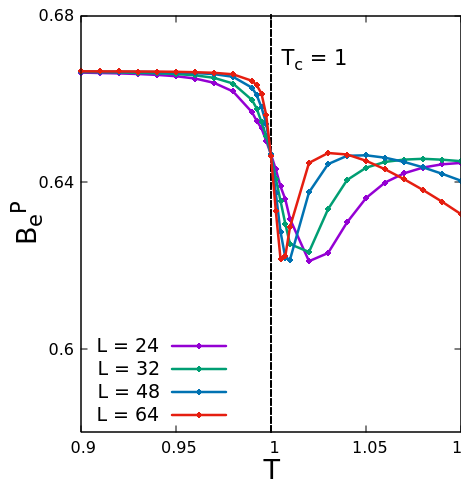}
    \subcaption{}
    \end{subfigure}
    \captionsetup{justification=raggedright, singlelinecheck=false}
    \caption{Plot of Binder cumulant $B_e^P$  vs. $T$  for different $L$ gives $T_c=1.$  (a) $\lambda = -0.2$, (b) $\lambda = -0.1$, (c) $\lambda = 0.0$, (d) $\lambda = 0.2.$    For each $\lambda,$ $J=J_c$  given by Eq. \eqref{eq:Jc} and data are averaged over $10^7$ samples.}
    \label{fig:supp2}
\end{figure}

\begin{figure}
    \centering
    \begin{subfigure}{0.245\linewidth}
    \includegraphics[width = \linewidth]{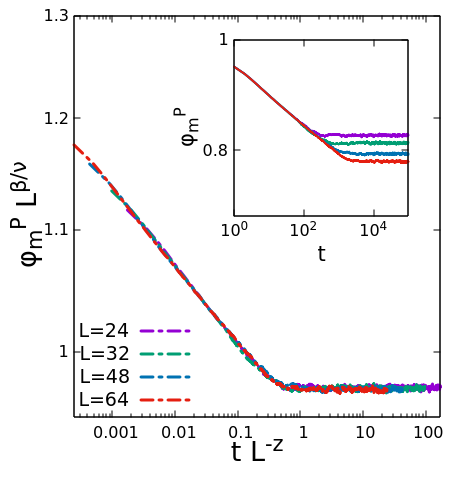}
    \subcaption{}
    \end{subfigure}
    \begin{subfigure}{0.245\linewidth}
    \includegraphics[width = \linewidth]{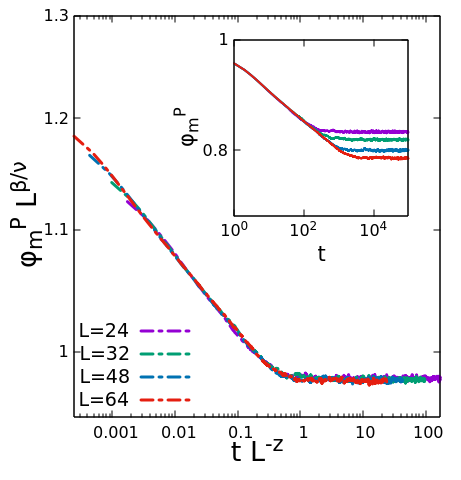}
    \subcaption{}
    \end{subfigure}
    \begin{subfigure}{0.245\linewidth}
    \includegraphics[width = \linewidth]{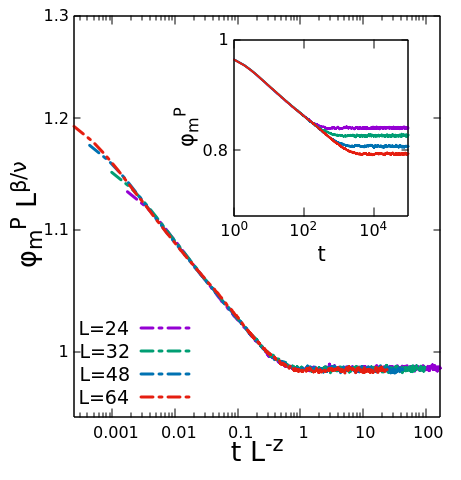}
    \subcaption{}
    \end{subfigure}
    \begin{subfigure}{0.245\linewidth}
    \includegraphics[width = \linewidth]{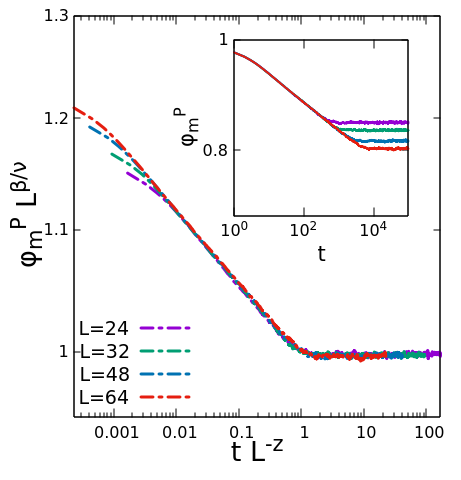}
    \subcaption{}
    \end{subfigure}
    \captionsetup{justification=raggedright, singlelinecheck=false}
        \caption{ Dynamical exponent $z$ and $\beta_m/\nu$ for magnetic percolation from the scaling collapse of $\phi_{m}^P L^{\frac{\beta_{m}^P}{\nu}}$ a function of $t/L^z,$ calculated at $T_c=1,$ $J_c$ given by \eqref{eq:Jc} for (a) $\lambda = -0.2$, (b) $\lambda = -0.1$, (c) $\lambda = 0.0$, (d) $\lambda = 0.2$. The raw data for different $L$ as a function of $t,$ averaged over $10^4$ runs are shown in the inset. The estimates of $z$ and $\beta_m/\nu$ are listed in Table \ref{table:III}.}  
    %\caption{ We rescale the x and y axis to obtain the collapsed temporal-data (magnetization) for (a) $\lambda = -0.2$, (b) $\lambda = -0.1$, (c) $\lambda = 0.0$, (d) $\lambda = 0.2$. Un-collapsed plots are in the inset.}
    \label{fig:supp3}
\end{figure}

\begin{figure}
    \centering
    \begin{subfigure}{0.245\linewidth}
\includegraphics[width = \linewidth]{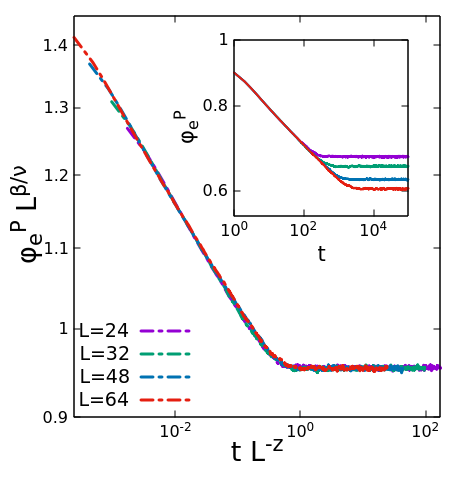}
    \subcaption{}
    \end{subfigure}
    \begin{subfigure}{0.245\linewidth}
    \includegraphics[width = \linewidth]{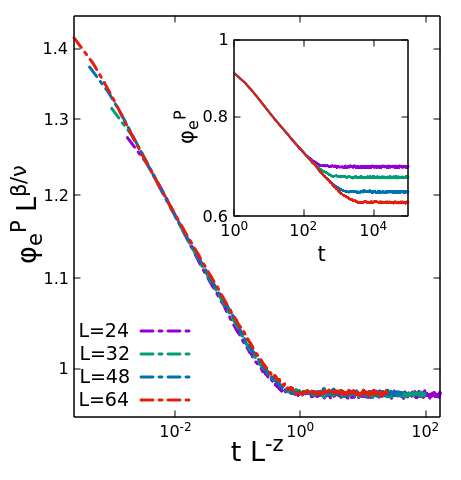}
    \subcaption{}
    \end{subfigure}
    \begin{subfigure}{0.245\linewidth}
    \includegraphics[width = \linewidth]{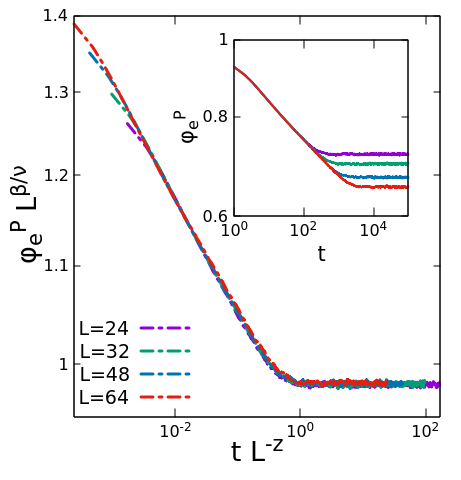}
    \subcaption{}
    \end{subfigure}
    \begin{subfigure}{0.245\linewidth}
    \includegraphics[width = \linewidth]{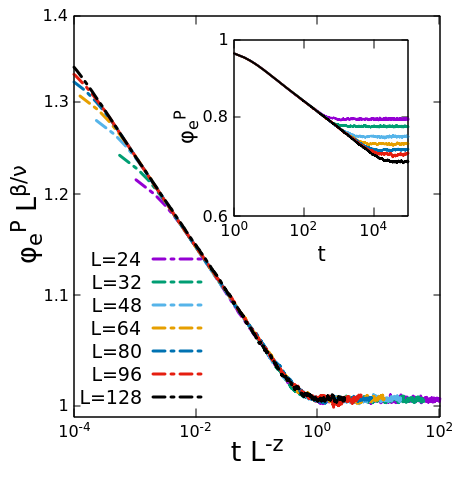}
    \subcaption{}
    \end{subfigure}
    \captionsetup{justification=raggedright, singlelinecheck=false}
    \caption{In a similar way, as in Fig. \ref{fig:supp3} here we estimate $z$ and $\beta_e/\nu$ for electric percolation
for (a) $\lambda = -0.2$, (b) $\lambda = -0.1$, (c) $\lambda = 0.0$, (d) $\lambda = 0.2$ and listed them in Table \ref{table:IV}.} 
    \label{fig:supp4}
\end{figure}

\begin{figure}
    \centering
    \begin{subfigure}{0.245\linewidth}
    \includegraphics[width = \linewidth]{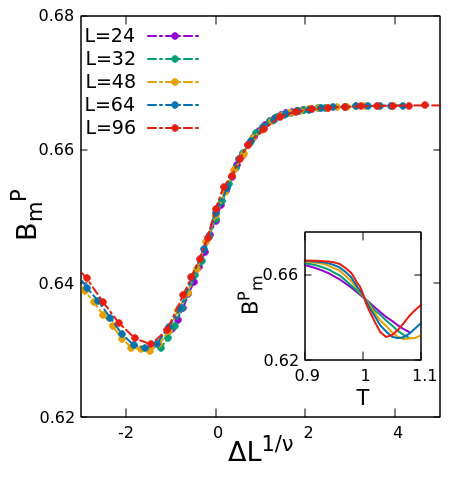}
    \subcaption{}
    \end{subfigure}
    \begin{subfigure}{0.245\linewidth}
    \includegraphics[width = \linewidth]{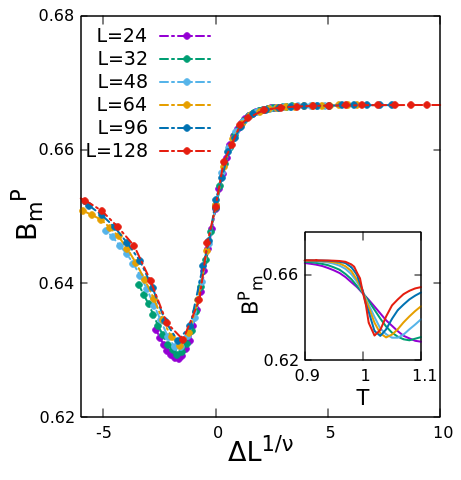}
    \subcaption{}
    \end{subfigure}
    \begin{subfigure}{0.245\linewidth}
    \includegraphics[width = \linewidth]{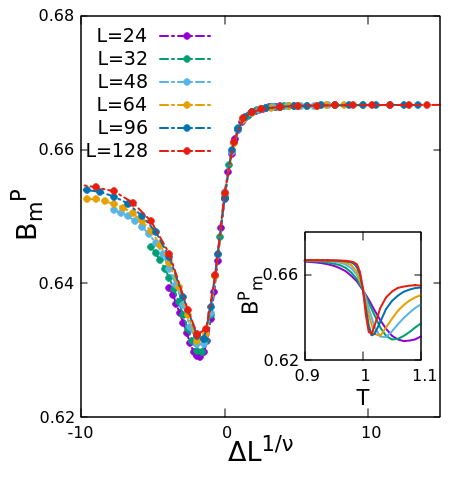}
    \subcaption{}
    \end{subfigure}
    \begin{subfigure}{0.245\linewidth}
    \includegraphics[width = \linewidth]{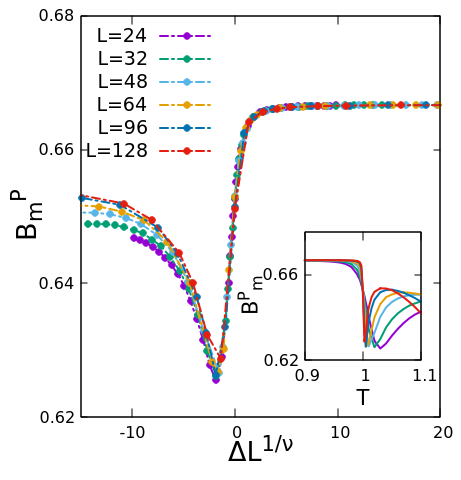}
    \subcaption{}
    \end{subfigure}
      \captionsetup{justification=raggedright, singlelinecheck=false}  
    \caption{Data collapse obtained for $B_m^P$ for different values of $\lambda$, (a) $\lambda = -0.2$, (b) $\lambda = -0.1$, (c) $\lambda = 0.0$, (d) $\lambda = 0.2$. Un-collapsed plots are in the inset. We average over $10^7$ samples taken from the steady state. The estimated critical exponents obtained from best collapse are listed in Table \ref{table:III}}
    \label{fig:supp5}
\end{figure}

\begin{figure}
    \centering
    \begin{subfigure}{0.245\linewidth}
    \includegraphics[width = \linewidth]{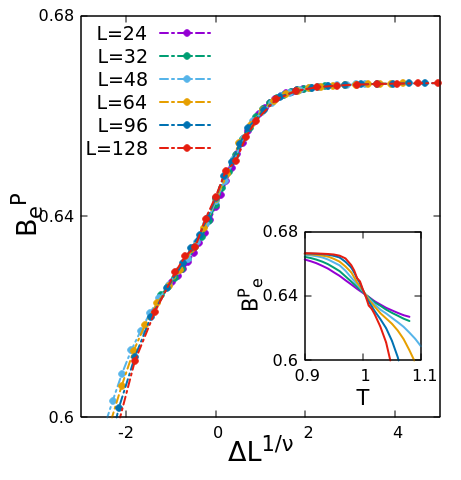}
    \subcaption{}
    \end{subfigure}
    \begin{subfigure}{0.245\linewidth}
    \includegraphics[width = \linewidth]{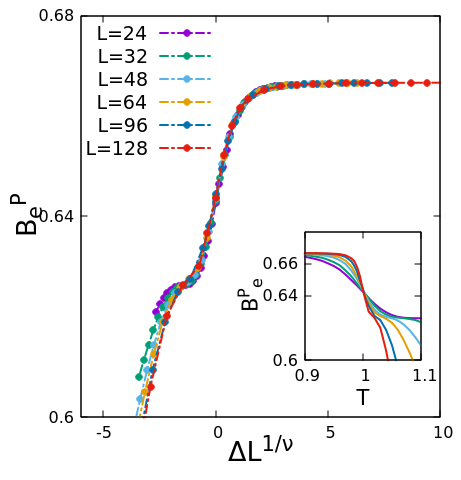}
    \subcaption{}
    \end{subfigure}
    \begin{subfigure}{0.245\linewidth}
    \includegraphics[width = \linewidth]{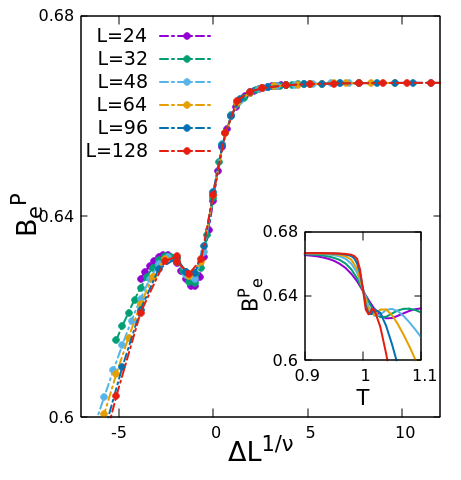}
    \subcaption{}
    \end{subfigure}
    \begin{subfigure}{0.245\linewidth}
    \includegraphics[width = \linewidth]{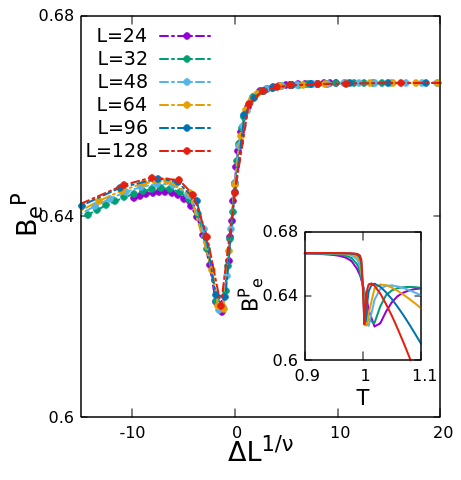}
    \subcaption{}
    \end{subfigure}
        \captionsetup{justification=raggedright, singlelinecheck=false}
    \caption{In a similar way as in Fig. \ref{fig:supp5} we plot the collapsed plots for $B_e^P$ for (a) $\lambda = -0.2$, (b) $\lambda = -0.1$, (c) $\lambda = 0.0$, (d) $\lambda = 0.2$ and listed the values of the critical exponents in Table \ref{table:IV}}
    \label{fig:supp6}
\end{figure}

\begin{figure}
    \centering
    \begin{subfigure}{0.245\linewidth}
    \includegraphics[width = \linewidth]{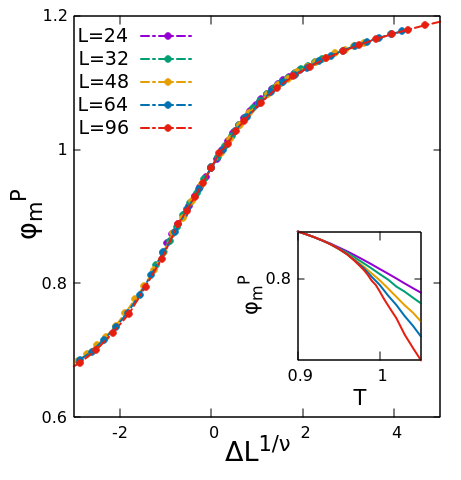}
    \subcaption{}
    \end{subfigure}
    \begin{subfigure}{0.245\linewidth}
    \includegraphics[width = \linewidth]{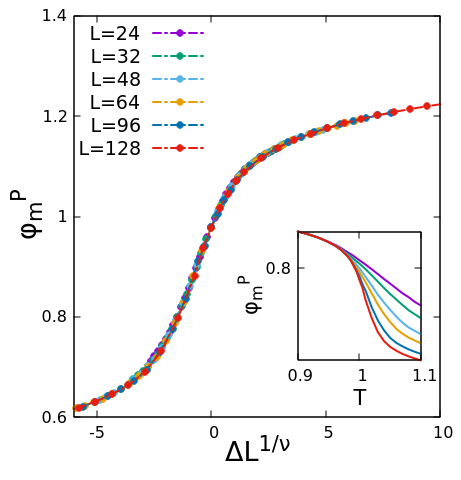}
    \subcaption{}
    \end{subfigure}
    \begin{subfigure}{0.245\linewidth}
    \includegraphics[width = \linewidth]{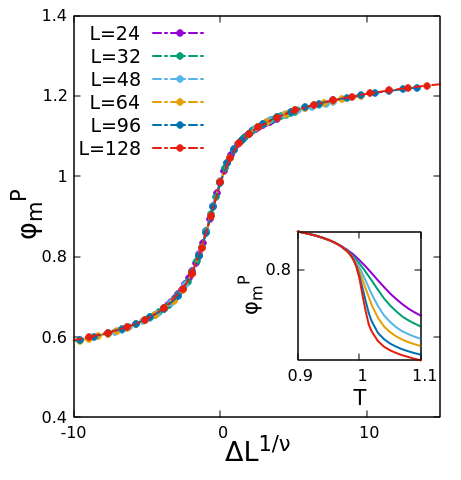}
    \subcaption{}
    \end{subfigure}
    \begin{subfigure}{0.245\linewidth}
    \includegraphics[width = \linewidth]{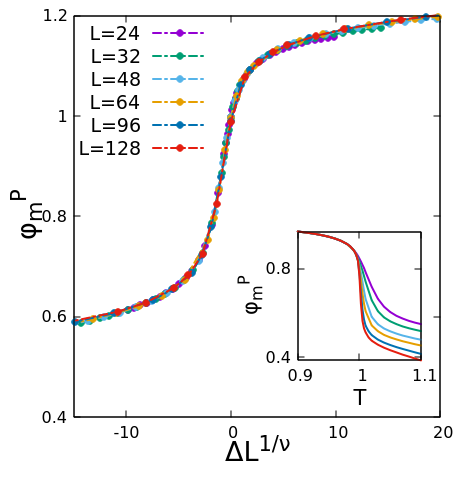}
    \subcaption{}
    \end{subfigure}   
    \captionsetup{justification=raggedright, singlelinecheck=false}
    \caption{Data collapse obtained with different system sizes for $\phi_m^P$ at values of $\lambda$, (a) $\lambda = -0.2$, (b) $\lambda = -0.1$, (c) $\lambda = 0.0$, (d) $\lambda = 0.2$. Un-collapsed plots are in the inset. We average over $10^7$ samples taken from the steady state. The estimated critical exponents obtained from best collapse are listed in Table \ref{table:III}}
    \label{fig:supp7}
\end{figure}

\begin{figure}
    \centering
    \begin{subfigure}{0.245\linewidth}
    \includegraphics[width = \linewidth]{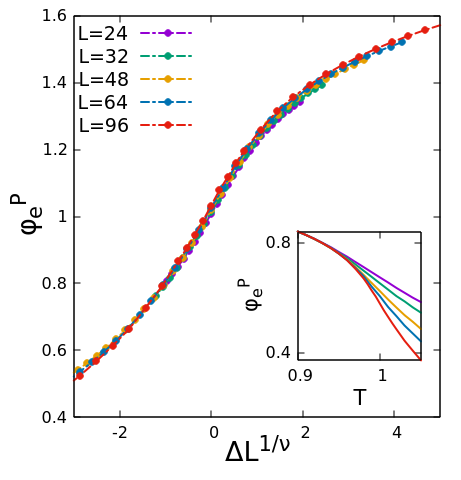}
    \subcaption{}
    \end{subfigure}
    \begin{subfigure}{0.245\linewidth}
    \includegraphics[width = \linewidth]{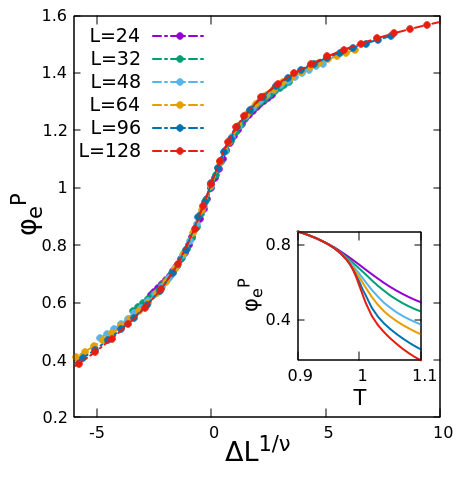}
    \subcaption{}
    \end{subfigure}
    \begin{subfigure}{0.245\linewidth}
    \includegraphics[width = \linewidth]{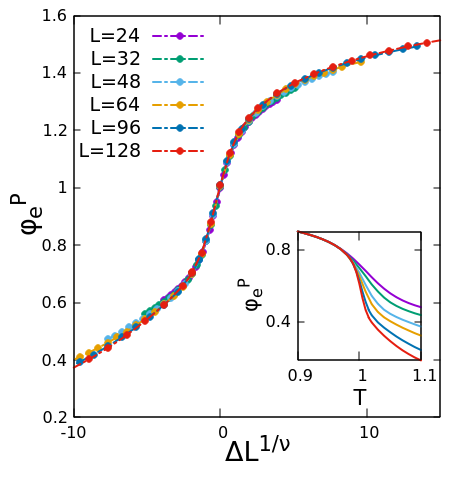}
    \subcaption{}
    \end{subfigure}
    \begin{subfigure}{0.245\linewidth}
    \includegraphics[width = \linewidth]{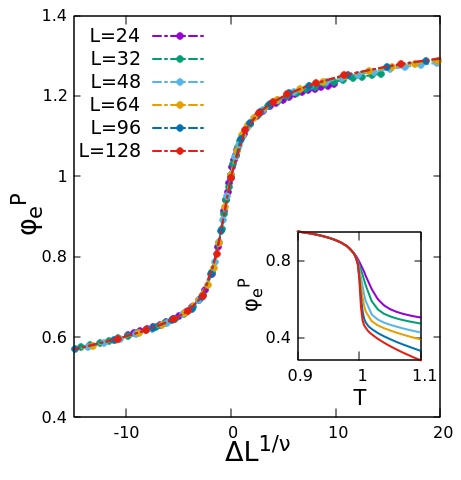}
    \subcaption{}
    \end{subfigure}  
    \captionsetup{justification=raggedright, singlelinecheck=false}
    \caption{In a similar way as in Fig. \ref{fig:supp7} we plot the collapsed plots for $\phi_e^P$ for (a) $\lambda = -0.2$, (b) $\lambda = -0.1$, (c) $\lambda = 0.0$, (d) $\lambda = 0.2$ and listed the values of the critical exponents in Table \ref{table:IV}}
    \label{fig:supp8}
\end{figure}

\begin{figure}
    \centering
    \begin{subfigure}{0.245\linewidth}
    \includegraphics[width = \linewidth]{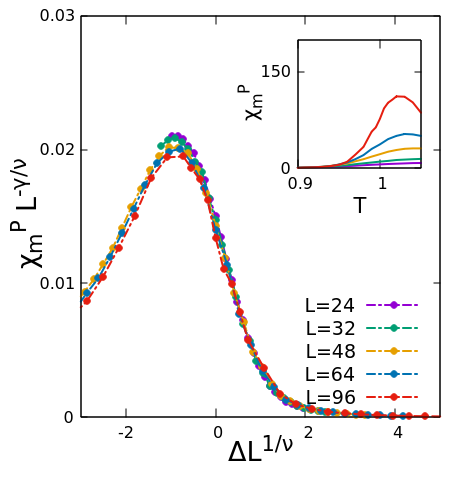}
    \subcaption{}
    \end{subfigure}
    \begin{subfigure}{0.245\linewidth}
    \includegraphics[width = \linewidth]{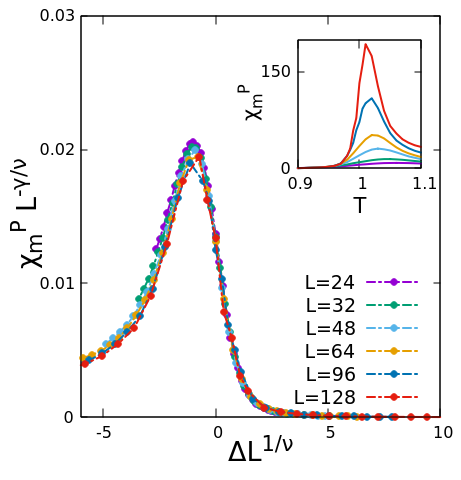}
    \subcaption{}
    \end{subfigure}
    \begin{subfigure}{0.245\linewidth}
    \includegraphics[width = \linewidth]{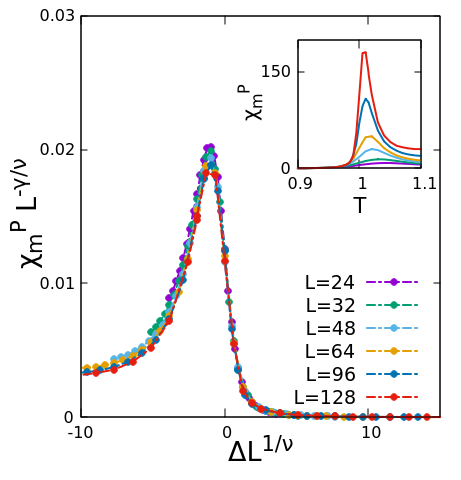}
    \subcaption{}
    \end{subfigure}
    \begin{subfigure}{0.245\linewidth}
    \includegraphics[width = \linewidth]{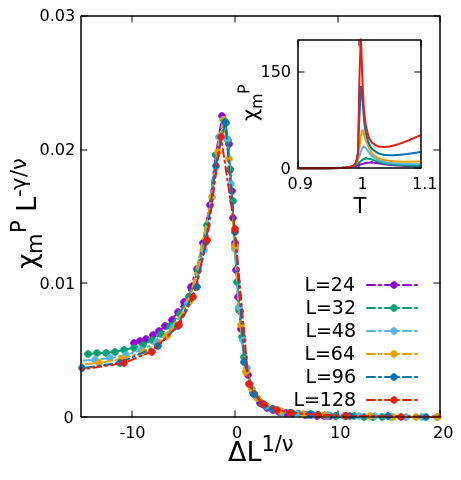}
    \subcaption{}
    \end{subfigure}
        \captionsetup{justification=raggedright, singlelinecheck=false}
    \caption{Data collapse obtained with different system sizes for $\chi_m^P$ at values of $\lambda$, (a) $\lambda = -0.2$, (b) $\lambda = -0.1$, (c) $\lambda = 0.0$, (d) $\lambda = 0.2$. Un-collapsed plots are in the inset. We average over $10^7$ samples taken from the steady state. The estimated critical exponents obtained from best collapse are listed in Table \ref{table:III}}
    \label{fig:supp9}
\end{figure}

\begin{figure}
    \centering
    \begin{subfigure}{0.245\linewidth}
    \includegraphics[width = \linewidth]{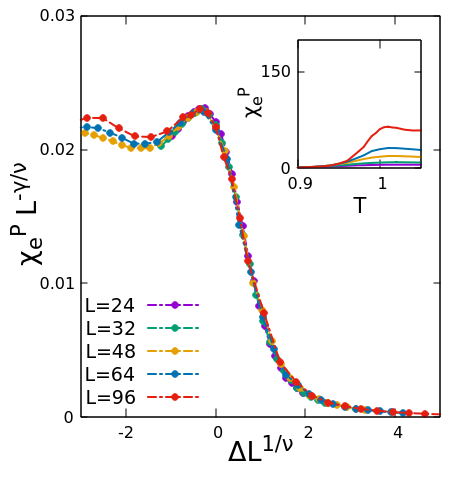}
    \subcaption{}
    \end{subfigure}
    \begin{subfigure}{0.245\linewidth}
    \includegraphics[width = \linewidth]{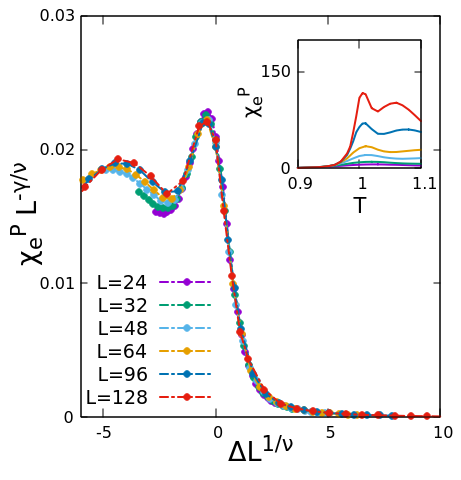}
    \subcaption{}
    \end{subfigure}
    \begin{subfigure}{0.245\linewidth}
    \includegraphics[width = \linewidth]{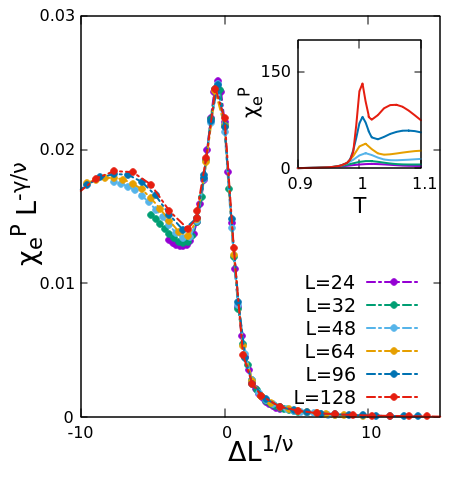}
    \subcaption{}
    \end{subfigure}
    \begin{subfigure}{0.245\linewidth}
    \includegraphics[width = \linewidth]{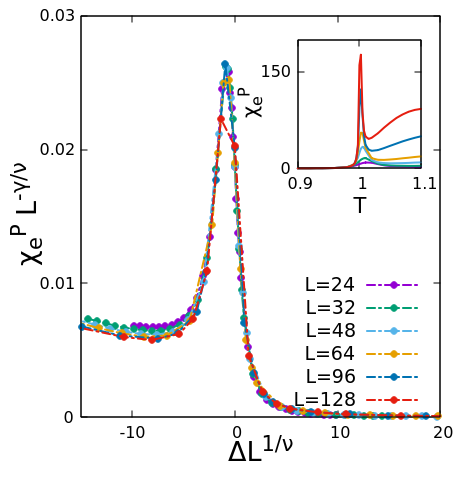}
    \subcaption{}
    \end{subfigure}
        \captionsetup{justification=raggedright, singlelinecheck=false}
    \caption{In a similar way as in Fig. \ref{fig:supp9} we plot the collapsed plots for $\phi_e^P$ for (a) $\lambda = -0.2$, (b) $\lambda = -0.1$, (c) $\lambda = 0.0$, (d) $\lambda = 0.2$ and listed the values of the critical exponents in Table \ref{table:IV}}
    \label{fig:supp10}
\end{figure}

\begin{figure}
    \centering
    \begin{subfigure}{0.245\linewidth}
    \includegraphics[width = \linewidth]{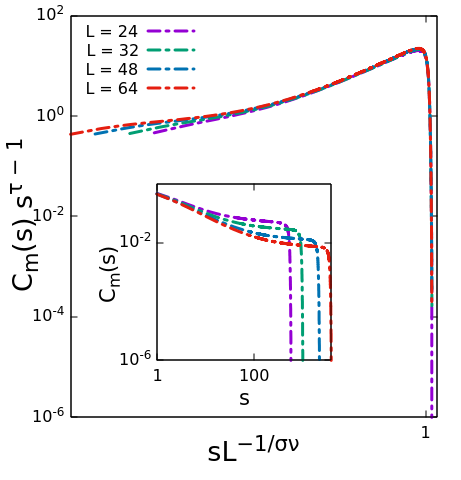}
    \subcaption{}
    \end{subfigure}
    \begin{subfigure}{0.245\linewidth}
    \includegraphics[width = \linewidth]{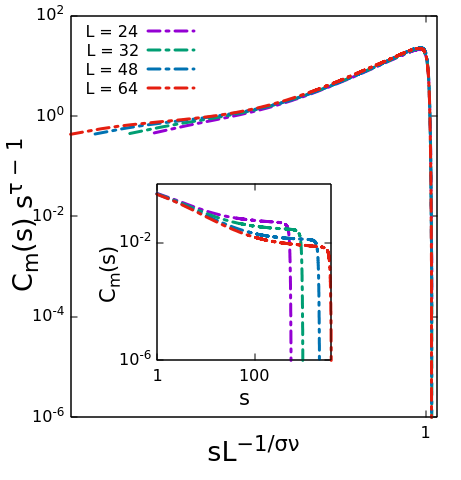}
    \subcaption{}
    \end{subfigure}
    \begin{subfigure}{0.245\linewidth}
    \includegraphics[width = \linewidth]{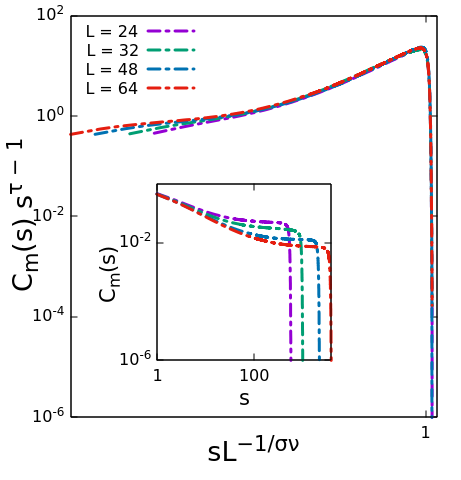}
    \subcaption{}
    \end{subfigure}
      \captionsetup{justification=raggedright, singlelinecheck=false}  
    \caption{Data collapse obtained with different sizes (L) from cluster (magnetization) size distribution for (a) $\lambda = -0.2$, (b) $\lambda = 0.0$, (c) $\lambda = 0.2$. Un-collapsed plots in the inset. The estimated critical exponents are reported in Table \ref{table:III}. Here we averaged over $10^5$ samples in steady state.}
    \label{fig:supp11}
\end{figure}

\begin{figure}
    \centering
    \begin{subfigure}{0.245\linewidth}
    \includegraphics[width = \linewidth]{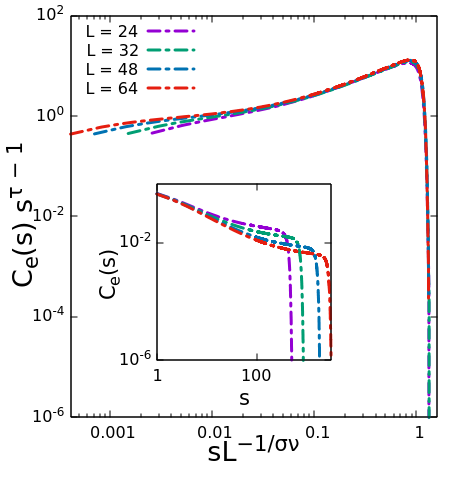}
    \subcaption{}
    \end{subfigure}
    \begin{subfigure}{0.245\linewidth}
    \includegraphics[width = \linewidth]{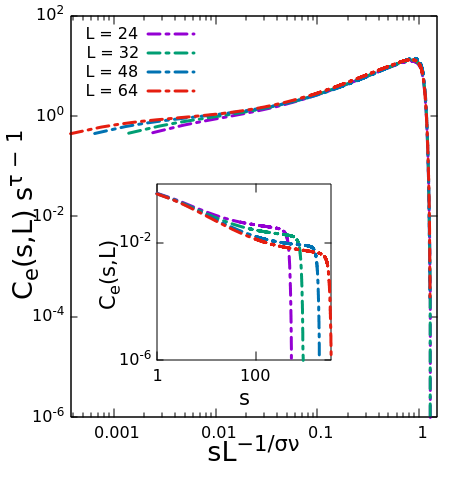}
    \subcaption{}
    \end{subfigure}
    \begin{subfigure}{0.245\linewidth}
    \includegraphics[width = \linewidth]{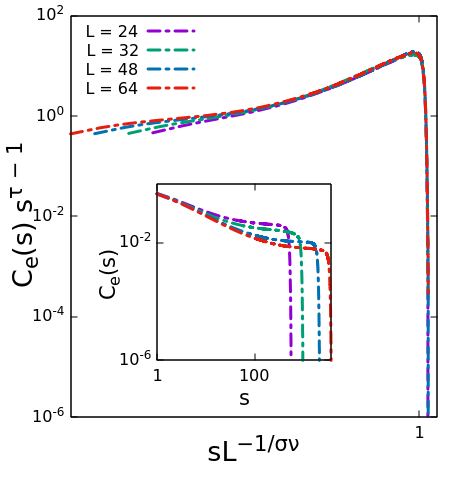}
    \subcaption{}
    \end{subfigure}
        \captionsetup{justification=raggedright, singlelinecheck=false}
    \caption{Similar plot as Fig. \ref{fig:supp11} for polarization cluster size distribution at (a) $\lambda = -0.2$, (b) $\lambda = 0.0$, (c) $\lambda = 0.2$. The estimated critical exponents are listed in Table \ref{table:IV}}
    \label{fig:supp12}
\end{figure}

\begin{table}[t]
\caption{ Critical exponents of magnetic percolation \label{table:III}}
\begin{tabular}{|c|c|c|c|c|c|c|c|c|c|c|} 
 \hline
$\lambda$ & $z$ &$\theta_m$& $\nu$ &$\beta_m$ & $\gamma_m$ &$\beta_m^P$ &$\gamma_m^P$ &$\tau_m$ &$\sigma_m$ &$D_m^P$\\ \hline
-0.2& 2.0(0) & 0.02(6)& 1.27(5) & 0.159&2.237 &0.06(6) &2.4(2) &2.0(3) &0.40(3)& 1.947\\ 
-0.1& 2.0(0) &0.02(6)&1.13(3) &0.142 &1.995 &0.05(9) &2.14(3) &2.0(3) &0.45(3) & 1.948\\ 
 0.0& 2.0(0) &0.02(6)&1.00(0) &0.125 &1.750 &0.052(1) &1.89(6) &2.0(3)& 0.51(3) & 1.948\\ 
 0.1& 2.0(0) &0.02(6)&0.87(8) &0.109 &1.541 &0.04(6) &1.66(3) &2.0(3) &0.58(5) & 1.948\\ 
 0.2& 2.0(1) &0.02(6)&0.77(1) &0.096 &1.372 &0.04(0) &1.46(2) &2.0(3) &0.66(6) &1.948\\ 
\hline
\end{tabular}
\end{table}
%%%%%%%%%%%%%%%%%%%%%%%%%%%%%%%%%%%
\begin{table}[h!]\caption{ Critical exponents of electric percolation \label{table:IV}}
\begin{tabular}{|c|c|c|c|c|c|c|c|c|c|c|} 
 \hline
$\lambda$ & $z$ &$\theta_e$& $\nu$ &$\beta_e$ & $\gamma_e$ &$\beta_e^P$ &$\gamma_e^P$ &$\tau_e^P$ & $\sigma_e$&$D_e^P$\\ \hline
-0.2& 2.0(0) & 0.06(0) &1.27(5) &0.388 &1.775 &0.16(0) &2.22(0) &2.0(8)& 0.41(9)&1.87\\
-0.1& 2.0(0) & 0.05(5) &1.13(3) &0.316 &1.634 &0.13(2) &2.00(2) &2.0(6) &0.46(8)& 1.88\\ 
 0.0& 2.0(0) & 0.05(0) &1.00(0) &0.250 &1.500 &0.10(4) &1.78(0) &2.0(5) &0.52(7) &1.90\\ 
 0.1& 2.0(5) &0.04(3)&0.87(8) &0.189 &1.380 &0.07(9) &1.58(8) &2.0(5) &0.59(6)&1.91\\
 0.2& 2.1(6) & 0.03(5) &0.77(1) &0.135 &1.272 &0.05(7) &1.42(6) &2.0(4)& 0.67(3) &1.926\\ 
\hline
\end{tabular}
\end{table}